\documentclass[pre,twocolumn,showpacs,superscriptaddress,preprintnumbers,floatfix]{revtex4-1}

\usepackage{etex}
\usepackage{ifpdf}
\usepackage{hyperref}
\usepackage{dcolumn}
\usepackage{url}
\usepackage{mathtools}
\usepackage{amsmath}
\usepackage{amscd}
\usepackage{amsfonts}
\usepackage{amssymb}
\usepackage{bm}   % bold math
\usepackage{bbm}
\usepackage{verbatim}
\usepackage{stmaryrd}
\usepackage{amsthm}
\usepackage{xcolor}
\usepackage{setspace}
\usepackage{tikz}
\usetikzlibrary{arrows}
\usetikzlibrary{automata}
\usetikzlibrary{positioning}
\usetikzlibrary{plotmarks}
\usepackage{pgfplots}
\pgfplotsset{compat=newest}

\tikzstyle{vaucanson}=[
  node distance=2.5cm,
  bend angle=15,
  auto,
  every loop/.style={},
  every edge/.style={->,draw=black,line width=1.2,>=latex,shorten <=1pt,shorten >=1pt},
  every state/.style={draw=black,line width=2,font=\large},
  loop right/.style={right,out=22,in=-22,loop},
  loop above/.style={above,out=112,in=68,loop},
  loop left/.style={left,out=202,in=158,loop},
  loop below/.style={below,out=292,in=248,loop},
  loop above right/.style={right,out=67,in=23,loop},
  loop above left/.style={left,out=157,in=113,loop},
  loop below left/.style={left,out=247,in=203,loop},
  loop below right/.style={right,out=337,in=293,loop},
]

\newcommand{\Symbol}[1]{\textcolor{blue}{#1}}
\newcommand{\Edge}[2]{$#1|\Symbol{#2}$}
\newcommand{\half}{\frac{1}{2}}

\theoremstyle{plain}    
\theoremstyle{plain}    
\theoremstyle{plain}    \newtheorem{Cor}{Corollary}
\theoremstyle{plain}    \newtheorem*{ProCor}{Proof}
\theoremstyle{plain}    \newtheorem{The}{Theorem}
\theoremstyle{plain}    \newtheorem*{ProThe}{Proof}
\theoremstyle{plain}    \newtheorem{Prop}{Proposition}
\theoremstyle{plain}    \newtheorem*{ProProp}{Proof}
\theoremstyle{plain}    
\theoremstyle{plain}    \newtheorem*{Rem}{Remark}
\theoremstyle{plain}    \newtheorem{Def}{Definition}
\theoremstyle{plain}    
\theoremstyle{plain}    \newtheorem{Alg}{Algorithm}

\newcommand{\eM}     {\mbox{$\epsilon$-machine}}
\newcommand{\eMs}    {\mbox{$\epsilon$-machines}}

\newcommand{\EMs}    {\mbox{$\epsilon$-Machines}}

\newcommand{\MeasAlphabet}  {\mathcal{A}}
\newcommand{\MeasSymbol}   { {X} }
\newcommand{\meassymbol}   { {x} }
\newcommand{\MeasSymbols}[2]{ \MeasSymbol_{#1:#2} }
\newcommand{\meassymbols}[2] { \meassymbol_{#1:#2} }

\newcommand{\Past} { \MeasSymbols{}{0} }
\newcommand{\Future} { \MeasSymbols{0}{} }
\newcommand{\future} { \meassymbols{0}{} }

\newcommand{\Present} {\MeasSymbol_0}

\newcommand{\CausalState}   { \mathcal{S} }

\newcommand{\causalstate}   { \sigma }
\newcommand{\CausalStateSet}    { \boldsymbol{\CausalState} }

\newcommand{\CausalEquivalence} { {\sim}_{\epsilon} }

\newcommand{\Prob}      {\Pr} % use standard command

\newcommand{\Cmu}       {C_\mu}
\newcommand{\hmu}       {h_\mu}
\newcommand{\EE}        {{\bf E}}

\newcommand{\SI}        {{\bf S}}
\newcommand{\SE}        { H_{\mathrm{sync}} }

\newcommand{\PC}        {\chi}

\newcommand{\forward}{+}
\newcommand{\reverse}{-}
\newcommand{\forwardreverse}{\pm} % \pm

\newcommand{\FutureCausalState} { {\CausalState}^{\forward} }

\newcommand{\PastCausalState}   { {\CausalState}^{\reverse} }

\newcommand{\lastindex}[2]{
  \edef\tempa{0}
  \edef\tempb{#2}
  \ifx\tempa\tempb
    \edef\tempc{#1}
  \else
    \edef\tempa{0}
    \edef\tempb{#1}
    \ifx\tempa\tempb
      \edef\tempc{#2}
    \else
      \edef\tempc{#1+#2}
    \fi
  \fi
  \tempc
}

\newcommand{\COrder}{k_{\PC}}

\newcommand{\MOrder}{R}
\newcommand{\SOrder}{k_{\SI}}
\newcommand{\Lsync} {\mathcal{L}_{\mathrm{sync}}}

\newcommand{\rmu}{r_\mu}
\newcommand{\bmu}{b_\mu}

\newcommand{\sigmamu}{\sigma_\mu}

\newcommand{\I}{\mathbf{I}}

\newcommand{\CSjoint}[1][,]{
   \edef\tempa{:}
   \edef\tempb{#1}
   \ifx\tempa\tempb
      \ensuremath{\FutureCausalState\!#1\PastCausalState}
   \else
      \ensuremath{\FutureCausalState#1\PastCausalState}
   \fi
}

\newif\ifpm
\edef\tempa{\forwardreverse}
\edef\tempb{\pm}
\ifx\tempa\tempb
   \pmfalse
\else
   \pmtrue
\fi

\renewcommand{\H}{\operatorname{H}}
\renewcommand{\I}{\operatorname{I}}

\newcommand{\edge}[1] {\stackrel{\Symbol{#1}}{\rightarrow}}

\colorlet {R_color}    {blue}
\colorlet {k_color}    {black!30!green}

\def\clap#1{\hbox to 0pt{\hss#1\hss}}

\def\mathclap{\mathpalette\mathclapinternal}

\def\mathclapinternal#1#2{%
\clap{$\mathsurround=0pt#1{#2}$}}

\begin{document}

\title{The Many Roads to Synchrony:\\
Natural Time Scales and Their Algorithms}

\author{Ryan G. James}
\email{rgjames@ucdavis.edu}
\affiliation{Complexity Sciences Center and Department of Physics, University of
  California at Davis, One Shields Avenue, Davis, CA 95616}

\author{John R. Mahoney}
\email{jmahoney3@ucmerced.edu}
\affiliation{School of Natural Sciences, University of California, Merced,
  California, 95344 }

\author{Christopher J. Ellison}
\email{cellison@cse.ucdavis.edu}
\affiliation{Complexity Sciences Center and Department of Physics, University of
  California at Davis, One Shields Avenue, Davis, CA 95616}

\author{James P. Crutchfield}
\email{chaos@ucdavis.edu}
\affiliation{Complexity Sciences Center and Department of Physics, University of
  California at Davis, One Shields Avenue, Davis, CA 95616}
\affiliation{Santa Fe Institute, 1399 Hyde Park Road, Santa Fe, NM
  87501}

\date{\today}
\bibliographystyle{unsrt}

% ************************* ABSTRACT *************************
\begin{abstract}
We consider two important time scales---the Markov and cryptic orders---that
monitor how an observer synchronizes to a finitary stochastic process. We show
how to compute these orders exactly and that they are most efficiently
calculated from the \eM, a process's minimal unifilar model. Surprisingly,
though the Markov order is a basic concept from stochastic process theory, it
is not a probabilistic property of a process. Rather, it is a topological
property and, moreover, it is not computable from any finite-state model other
than the \eM. Via an exhaustive survey, we close by demonstrating that infinite
Markov and infinite cryptic orders are a dominant feature in the space of
finite-memory processes. We draw out the roles played in statistical mechanical
spin systems by these two complementary length scales.

\vspace{0.1in}

\end{abstract}

\preprint{Santa Fe Institute Working Paper 10-11-025}
\preprint{arXiv:1010.5545 [nlin.CD]}

\maketitle
\
% ****************************************************************

%\tableofcontents
\setstretch{1.1}

\section{Introduction}
\label{sec:intro}

Stochastic processes are frequently characterized by the spatial and
temporal length scales over which correlations exist. In physics, the
range of correlations is a structural property giving, for example,
the distance over which significant energetic coupling exists among a
system's degrees of freedom \cite{Reic80a}. In time series analysis,
knowing the temporal scale of correlations is key to successful
forecasting \cite{Weig93a}. In biosequence analysis, the decay of
correlations along DNA base pairs determines in some measure the
difficulty faced by a replicating enzyme as it ``decides'' to begin
transcribing a gene \cite{Raj08a}. In multiagent systems, one of an
agent's first goals is to detect useful states in its environment
\cite{Crut01b}. The common element in these is that the correlation
scale determines how quickly an observer---analyst, forecaster,
enzyme, or agent---\emph{synchronizes} to a process; that is, how it
comes to know a relevant structure of the stochastic process.

We recently showed that there are a number of distinct, though
related, length scales associated with synchronizing to stationary
stochastic processes \cite{Crutchfield2010}. Here, we show that these
length scales are \emph{topological}, depending only on the underlying
graph topology of a canonical representation of the stochastic
process. This reveals deep ties between the structure of a process's
minimal sufficient statistic and synchronization of an observer. We
also recently introduced another class of synchronization length
scales based, not on state-based models, but on the convergence of
sequence statistics \cite{James2011}. We briefly compare these to the
Markov and cryptic orders.

Specifically, we investigate measures of synchronization and their associated
lengths scales for hidden Markov models (HMMs)---a particular class of
processes with an internal (hidden) Markovian dynamic that produces an observed
sequence. We focus on two such measures---the Markov order and the cryptic
order---and show through a series of incremental steps how they can be
efficiently and accurately computed from the process's minimal sufficient
statistic, the \eM.

Our development proceeds as follows. After briefly outlining the required
background in Sec. \ref{sec:defs}, we introduce the two primary measures of
interest in Sec.~\ref{sec:problem-statement} and demonstrate their calculation
via naive methods in Sec.~\ref{sec:naive-approach}. Reflecting on a surprising
finding in Sec.~\ref{sec:topological}, Sec.~\ref{sec:synchronizing-words} shows
to how alleviate several weaknesses in the naive approach. Then, borrowing
relevant data structures from formal language theory,
Sec.~\ref{sec:subset-construction} resolves the last of the issues.  Together
these steps provide an efficient algorithm for exactly calculating the Markov
order when it is finite and for determining when it is infinite.  Building on
this new understanding, Sec.~\ref{sec:cryptic-order} goes on to show how to
compute the second time scale---the cryptic order---through similar means. We
then briefly touch upon other time scales and their relative bounds in
Sec.~\ref{sec:other-orders}.  Leveraging the computational efficiency, we
survey the Markov and cryptic orders among \eMs\ in Sec.~\ref{sec:survey} and
conclude that infinite correlation is a dominate property in the space of
memoryful stationary processes. The implication is that most, if not all,
observers cannot synchronize exactly~\cite{Travers2010a}. To illustrate how these time scales apply in practice, Sec.~\ref{sec:spin-systems} characterizes
correlations in one-dimensional spin systems. Finally, we conclude by
discussing how these time scales compare to other measures of interest and by
suggesting applications where they and their algorithms will prove useful.

\section{Background}
\label{sec:defs}

We assume the reader has introductory knowledge of information theory
and finite-state machines, such as that found in the first few
chapters of Ref.~\cite{Cover2006} and Ref.~\cite{Hopcroft2001},
respectively. Our development makes particular use of \eMs, a natural
representation of a process that makes many properties directly and
easily calculable; for a review see Ref.~\cite{Crut12a}. A cursory
understanding of symbolic dynamics, such as found in the first few
chapters of Ref.~\cite{Lind1999} is useful for several results.

We denote subsequences in a time series as $\MeasSymbols{a}{b}$, where
$a < b$, to refer to the random variable sequence $\MeasSymbol_{a}
\MeasSymbol_{a+1} \MeasSymbol_{a+2} \cdots \MeasSymbol_{b-1}$, which
has length $b-a$. We drop an index when it is infinite. For example,
the \emph{past} $\MeasSymbols{-\infty}{0}$ is denoted $\Past$ and the
\emph{future} $\MeasSymbols{0}{\infty}$ is denoted $\Future$. We
generally use $w$ to refer to a \emph{word}---a sequence of symbols
drawn from an alphabet $\MeasAlphabet$. We place two words, $u$ and
$v$, adjacent to each other to mean concatenation: $w = uv$. We define
a \emph{process} to be a joint probability distribution over
$\MeasSymbols{}{} = \Past \Future$.

A \emph{presentation} of a given process is any state-based
representation that generates the process. A process's \emph{\eM} is
its unique, minimal unifilar presentation~\footnote{Unifilar is known
  as ``deterministic'' in the finite automata literature. Here, we avoid
  the latter term so that confusion does not arise due to the stochastic
  nature of the models being used. It is referred to as ``right-resolving'' in
  symbolic dynamics \cite{Lind1999}.}. The recurrent states of a
process's \eM\ are known as the \emph{causal states} and, at time $t$,
are denoted $\CausalState_t$. The causal states are the minimal
sufficient statistic of $\Past$ about $\Future$. For a thorough
treatment on presentations see Ref.~\cite{Crutchfield2010}.

\section{Problem Statement}
\label{sec:problem-statement}

When confronted with a process, one of the most natural questions to
ask is, How much memory does it have? Is it like a coin or a die,
with no memory? Does it alternate between two values, requiring that
the process remember its phase? Does it express patterns that are
arbitrarily long, requiring an equally long memory? This type of
memory is quantified by the Markov order:
\begin{align}
  \MOrder \equiv \min \left\{ \ell ~|~ \Prob(\Present |
    \MeasSymbols{-\ell}{0}) = \Prob(\Present | \Past) \right\}
  ~.
\label{eq:MarkovOrder}
\end{align}
To put it colloquially, how many prior observations must one remember
to predict as well as remembering the infinite past? Markov chains
have $\MOrder = 1$ by their very definition. Hidden Markov models,
though their internal dynamics are Markovian ($\MOrder = 1$), their
observed behavior can range from memoryless ($\MOrder = 0$) to
infinite ($\MOrder = \infty$). A major goal in the following is to
show how to compute a process's $\MOrder$ efficiently and accurately
given its \eM. In this vein it is prudent to recast
Eq.~(\ref{eq:MarkovOrder}) using causal states:
\begin{align}
\Prob(\Present | \MeasSymbols{-\MOrder}{0}) & = \Prob(\Present | \Past)
        \nonumber \\
  \implies& \Past \,\CausalEquivalence\, \MeasSymbols{-\MOrder}{0} \nonumber \\
  \implies& \H[ \CausalState_0 | \MeasSymbols{-\MOrder}{0} ] = 0 \nonumber \\
  \implies& \MOrder = \min \left\{ \ell ~|~ \H[\CausalState_0 |
            \MeasSymbols{-\ell}{0} ] = 0 \right\} \nonumber \\
          & \phantom{\MOrder} = \min \left\{ \ell ~|~ \H[\CausalState_\ell |
            \MeasSymbols{0}{\ell} ] = 0 \right\}
  ~.
\label{eq:MarkovOrdereM}
\end{align}
In effect, since the past $\MOrder$ observations predict just as well
as the infinite past, the causal states are a function of
length-$\MOrder$ pasts.

The second primary length scale we discuss is the \emph{cryptic order}
$\COrder$ \cite{Mahoney2009}. Its definition builds from Eq.
(\ref{eq:MarkovOrdereM}):
\begin{align}
  \COrder \equiv \min \left\{ \ell ~|~ \H[\CausalState_\ell | \Future ] = 0 \right\}
  ~.
\label{eq:CrypticOrder}
\end{align}
The difference between the two is that cryptic order is conditioned on
the \emph{infinite} future, as opposed to a finite one. This provides
our interpretation of the cryptic order: $\COrder$ is the number of
causal states that cannot be \emph{retrodicted}. That is, no matter
how many future symbols we know, the first $\COrder$ internal states
the process visited cannot be inferred.

\section{Naive Approach}
\label{sec:naive-approach}

To illustrate a direct method of determining a process's Markov and cryptic
orders, we appeal to yet another form of their
definitions~\cite{Crutchfield2010}:
\begin{align}
  \MOrder &= \min \left\{\ell ~|~ \H[\MeasSymbols{0}{\ell}] = \EE +
                                 \ell \, \hmu \right\} \\
  \COrder &= \min \left\{\ell ~|~ \H[\MeasSymbols{0}{\ell},
                                 \CausalState_\ell] = \EE + \ell \, \hmu
                                 \right\}
  ~,
\label{eq:OrdersLinearAsymptote}
\end{align}
where $\EE = \I[\Past, \Future]$ is known as the \emph{excess entropy}
and $\hmu = \H[\Present | \Past]$ is known as the \emph{entropy rate}
\cite{Crutchfield2008}. The intuition for these is identical to those
above: Once we reach Markov (cryptic) order, we predict as accurately
as possible. It is worth noting that these definitions only hold for
\emph{finitary} ($\EE < \infty$), stationary processes.

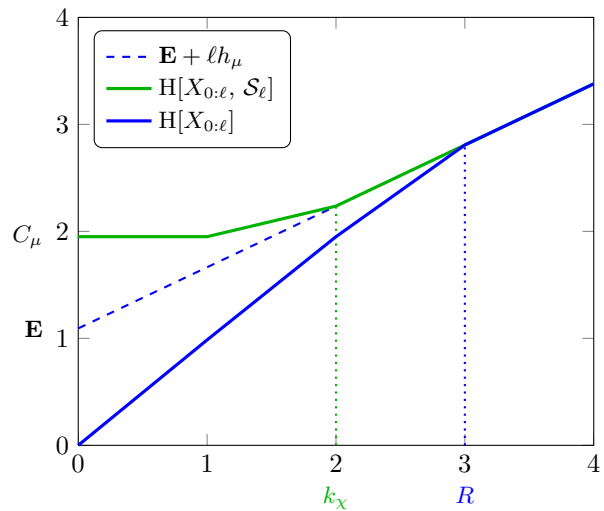
\begin{figure}
  \centering
  \begin{tikzpicture}
    \begin{axis}[
        clip=false,
        xmin=0,
        xmax=4,
        ymin=0,
        ymax=4,
        legend pos=north west,
        legend cell align=left,
        legend style={
          rounded corners=3pt,
          font=\small,
          inner sep=4pt,
        }
      ]
      \draw [thick, style=dotted, R_color] (axis cs: 3, 0) -- (axis cs: 3, 2.80735492);
      \draw [thick, style=dotted, k_color] (axis cs: 2, 0) -- (axis cs: 2, 2.23592635);
      \addplot [thick, blue, dashed] coordinates {
        (0, 1.0930692077718898)
        (4, 3.37878349)
      };
      \addplot [very thick, k_color] coordinates {
        (0, 1.95021206)
        (1, 1.95021206)
        (2, 2.23592635)
        (3, 2.80735492)
        (4, 3.37878349)
      };
      \addplot [very thick, R_color] coordinates {
        (0, 0)
        (1, 0.98522814)
        (2, 1.95021206)
        (3, 2.80735492)
        (4, 3.37878349)
      };
      \draw (axis cs: -0.2, 1.0930692077718898) node [anchor=east] {\small $\EE$};
      \draw (axis cs: -0.2, 1.9502120649147467) node [anchor=east] {\small $\Cmu$};
      \draw (axis cs: 3, -0.3) node [R_color, anchor=north] {\small $\MOrder$};
      \draw (axis cs: 2, -0.3) node [k_color, anchor=north] {\small $\COrder$};
      \legend{
        $\EE + \ell \hmu$,
        $\H[\MeasSymbols{0}{\ell}\text{, } \CausalState_\ell]$,
        $\H[\MeasSymbols{0}{\ell}]$
      };
    \end{axis}
  \end{tikzpicture}
  \caption{Block entropy and block-state entropy for the PSB Process
    of Fig. \ref{fig:machine}: The block entropy curve reaches its
    asymptotic behavior ($\EE + \ell \, \hmu$) at $\ell = 3$,
    indicating a Markov order $R = 3$. The block-state entropy curve
    reaches the same asymptote at $\ell = 2$ and so the process is
    cryptic order $\COrder = 2$. }
\label{fig:entropy-curves}
\end{figure}

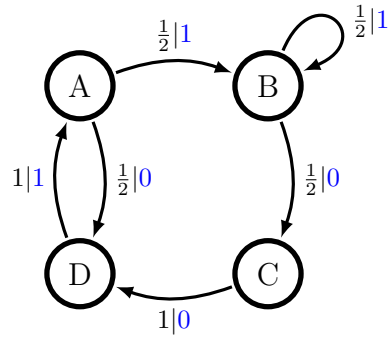
\begin{figure}
  \centering
  \begin{tikzpicture}[style=vaucanson, bend angle=20]
    \node [state] (A)              {A};
    \node [state] (B) [right of=A] {B};
    \node [state] (C) [below of=B] {C};
    \node [state] (D) [below of=A] {D};

    \path (A) edge [bend left]        node { \Edge{\half}{1}} (B)
          (A) edge [bend left]        node { \Edge{\half}{0}} (D)
          (B) edge [loop above right] node {~\Edge{\half}{1}} (B)
          (B) edge [bend left]        node { \Edge{\half}{0}} (C)
          (C) edge [bend left]        node { \Edge{1}{0}}     (D)
          (D) edge [bend left]        node { \Edge{1}{1}}     (A);
  \end{tikzpicture}
  \caption{The Phase-Slip Backtrack (PSB) Process: Edges are labeled
    \Edge{p}{s} where $p$ is the probability of an edge being followed
    and $s$ is the symbol emitted upon traversing it. }
\label{fig:machine}
\end{figure}

These definitions lead to a simple way of determining a process's
Markov and cryptic orders. To compute the Markov order, we calculate
the entropy $\H[\MeasSymbols{0}{\ell}]$ of longer and longer blocks of
contiguous observations until it begins to grow linearly. We call this
function of $\ell$ the \emph{block entropy curve}. The first $\ell$ at
which $\H[\MeasSymbols{0}{\ell}]$ matches its linear asymptote is the
Markov order. To compute the cryptic order, we perform a similar test,
but rather than calculating the entropy of blocks of observations
alone, we calculate the entropy
$\H[\MeasSymbols{0}{\ell},\CausalState_\ell]$ of those blocks along
with the causal states that are induced by those observations. We call this function of $\ell$ the \emph{block-state entropy curve}.  The cryptic order is the length at which the block-state entropy curve reaches its asymptotic linear behavior. This view of the two orders is shown in Fig.~\ref{fig:entropy-curves}. The data for the block entropy and block-state entropy curves shown there comes from the \emph{Phase-Slip Backtrack} (PSB) Process show in
Fig.~\ref{fig:machine}.

It is important to point out the weaknesses of this approach. They are
at least fourfold, one must (i) know $\hmu$ exactly, (ii) know $\EE$
exactly, (iii) be able to differentiate the block entropies being
\emph{exactly on} the asymptote from \emph{less than machine precision
  away from} the asymptote, and (iv) be able to ``guess'' when
$\MOrder$ or $\COrder$ are infinite in order to terminate the
calculation. The first two are not prohibitive. The entropy rate
$\hmu$ can be computed exactly from any unifilar model of the process,
and so its calculation can be done fairly easily~\cite{Shalizi2008}.
Similarly, the excess entropy $\EE$ can be computed if the joint
distribution over both a unifilar, gauge-free model of the process and
a unifilar, gauge-free model of the reverse of the process is on
hand~\cite{Ellison2009}.

The last two weaknesses do not have such direct solutions. How are we
to know if our entropy calculation at length $\ell$ is exactly equal
to $\EE + \ell \hmu$? Or, instead, are the curve and linear asymptote
so close that finite-precision estimates cannot differentiate them?
Compounding this, what if $\H[\MeasSymbols{0}{\ell}]$ has not equaled
$\EE + \ell \hmu$ by $\ell = 10^6$? Can one assume that it ever will?
Perhaps the process is Markov order $R = 10^8$. These are the two
particular weaknesses that need to be overcome.

\section{Markov Order is Topological}
\label{sec:topological}

We start with the somewhat surprising observation that Markov order is
not a probabilistic property, as seemingly suggested by Eq.
(\ref{eq:MarkovOrder}), but rather a topological one. The first hint
at this comes, though, in an empirical study. The question then
becomes just how is this so. By way of answering it, we solve the
fundamental problems noted with the naive approach to Markov order.
Several examples serve to drive home the idea and illustrate the
calculation methods.

\subsection{An Observation}

\begin{figure}
  \centering
  \begin{tikzpicture}[style=vaucanson, bend angle=20]
    \node [state] (A)              {A};
    \node [state] (B) [right of=A] {B};
    \node [state] (C) [below of=B] {C};
    \node [state] (D) [below of=A] {D};

    \path (A) edge [bend left]        node { \Edge{1-p}{1}} (B)
          (A) edge [bend left]        node { \Edge{p}{0}}   (D)
          (B) edge [loop above right] node {~\Edge{1-q}{1}} (B)
          (B) edge [bend left]        node { \Edge{q}{0}}   (C)
          (C) edge [bend left]        node { \Edge{1}{0}}   (D)
          (D) edge [bend left]        node { \Edge{1}{1}}   (A);
  \end{tikzpicture}
\caption{Phase-Slip Backtrack Process with parametrized transition
  probabilities.
  }
\label{fig:machine-topological}
\end{figure}
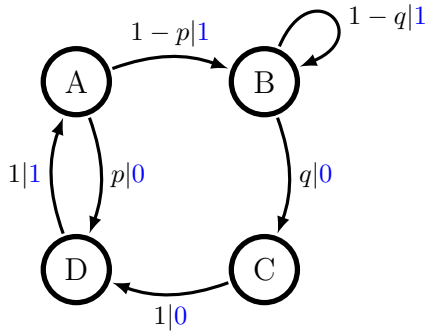

\begin{figure}
  \centering
  \includegraphics[width=\columnwidth]{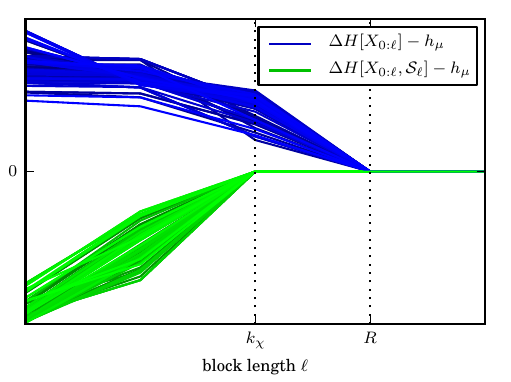}
  \caption{Entropy convergence curves versus block length $\ell$ for
    Fig.~\ref{fig:machine-topological}'s family of processes with
    several dozen random values for $p$ and $q$. The linear asymptotic
    behavior ($\hmu$) has been subtracted out of each curve. (See
    inset.) The Markov order $\MOrder$ and cryptic order $\COrder$ are
    the lengths $\ell$ at which the blue (darker) and green (lighter)
    lines, respectively, reach zero. Thus, both orders are independent
    of the generating machine's probability parameters. }
\label{fig:family}
\end{figure}

The first step forward in solving the two main problems encountered in the
naive Markov order method is to take a step back. Rather than considering the
particular process generated by the machine in Fig.~\ref{fig:machine}, we study
the family of processes generated when its transition probabilities are varied
while the structure remains the same. This family is shown by the parametrized
machine of Fig.~\ref{fig:machine-topological}. If we compute block and
block-state entropy curves for a random ensemble of processes from this family,
plot the derivative of those curves and subtracting out their asymptotic
behavior, we arrive at the block and block-state entropy convergence shown in
Fig.~\ref{fig:family}.

As it dramatically demonstrates, the Markov and cryptic orders are
\emph{independent} of the transition probabilities in the machine's
structure. Thus, any pattern relevant for prediction is encoded by the
\eM's topology.

\subsection{Synchronizing Words}
\label{sec:synchronizing-words}

On careful inspection of Eq.~(\ref{eq:MarkovOrdereM}), however, it is
not surprising that the Markov order is a topological property. A
conditional entropy $\H[X | Y]$ vanishes only if $X$ is a
deterministic function of $Y$. In our case, $\H[\CausalState_\MOrder |
\MeasSymbols{0}{\MOrder}] = 0$ means that each length-$\MOrder$ word
determines a unique state of the model. We say that each word of
length $\MOrder$ is \emph{synchronizing} \cite{Crutchfield2010}.
(Later, we consider only \emph{prefix-free} synchronizing
words---those which have no initial subword that also synchronizes.)
If one observes a process having no inkling as to which state its
hidden Markov model began in, then after observing $\MOrder$ symbols
the exact state will be known.

\begin{figure}
  \centering
  \begin{tikzpicture}
    % \draw [help lines] (0, 0) grid (7, 8);

    % 001
    \draw (0.5, 6.5) node {\Symbol{0}};
    \draw (1.5, 6.5) node {\Symbol{0}};
    \draw (2.5, 6.5) node {\Symbol{1}};

    \draw (0, 7.75) node {\small A};
    \draw (0, 7.50) node {\small B};
    \draw (0, 7.25) node {\small C};

    \draw (0.15, 7.75) -- (0.85, 7.00);
    \draw (0.15, 7.50) -- (0.85, 7.25);
    \draw (0.15, 7.25) -- (0.85, 7.00);

    \draw (1, 7.25) node {\small C};
    \draw (1, 7.00) node {\small D};

    \draw (1.15, 7.25) -- (1.85, 7.00);

    \draw (2, 7.00) node {\small D};

    \draw (2.15, 7.00) -- (2.85, 7.75);

    \draw (3, 7.75) node {\small A};

    % 010
    \draw (0.5, 4.5) node {\Symbol{0}};
    \draw (1.5, 4.5) node {\Symbol{1}};
    \draw (2.5, 4.5) node {\Symbol{0}};

    \draw (0, 5.75) node {\small A};
    \draw (0, 5.50) node {\small B};
    \draw (0, 5.25) node {\small C};

    \draw (0.15, 5.75) -- (0.85, 5.00);
    \draw (0.15, 5.50) -- (0.85, 5.25);
    \draw (0.15, 5.25) -- (0.85, 5.00);

    \draw (1, 5.25) node {\small C};
    \draw (1, 5.00) node {\small D};

    \draw (1.15, 5.00) -- (1.85, 5.75);

    \draw (2, 5.75) node {\small A};

    \draw (2.15, 5.75) -- (2.85, 5.00);

    \draw (3, 5.00) node {\small D};

    % 011
    \draw (0.5, 2.5) node {\Symbol{0}};
    \draw (1.5, 2.5) node {\Symbol{1}};
    \draw (2.5, 2.5) node {\Symbol{1}};

    \draw (0, 3.75) node {\small A};
    \draw (0, 3.50) node {\small B};
    \draw (0, 3.25) node {\small C};

    \draw (0.15, 3.75) -- (0.85, 3.00);
    \draw (0.15, 3.50) -- (0.85, 3.25);
    \draw (0.15, 3.25) -- (0.85, 3.00);

    \draw (1, 3.25) node {\small C};
    \draw (1, 3.00) node {\small D};

    \draw (1.15, 3.00) -- (1.85, 3.75);

    \draw (2, 3.75) node {\small A};

    \draw (2.15, 3.75) -- (2.85, 3.50);

    \draw (3, 3.50) node {\small B};

    % 100
    \draw (0.5, 0.5) node {\Symbol{1}};
    \draw (1.5, 0.5) node {\Symbol{0}};
    \draw (2.5, 0.5) node {\Symbol{0}};

    \draw (0, 1.75) node {\small A};
    \draw (0, 1.50) node {\small B};
    \draw (0, 1.00) node {\small D};

    \draw (0.15, 1.75) -- (0.85, 1.50);
    \draw (0.15, 1.50) -- (0.85, 1.50);
    \draw (0.15, 1.00) -- (0.85, 1.75);

    \draw (1, 1.75) node {\small A};
    \draw (1, 1.50) node {\small B};

    \draw (1.15, 1.75) -- (1.85, 1.00);
    \draw (1.15, 1.50) -- (1.85, 1.25);

    \draw (2, 1.25) node {\small C};
    \draw (2, 1.00) node {\small D};

    \draw (2.15, 1.25) -- (2.85, 1.00);

    \draw (3, 1.00) node {\small D};

    % 101
    \draw (4.5, 5.5) node {\Symbol{1}};
    \draw (5.5, 5.5) node {\Symbol{0}};
    \draw (6.5, 5.5) node {\Symbol{1}};

    \draw (4, 6.75) node {\small A};
    \draw (4, 6.50) node {\small B};
    \draw (4, 6.00) node {\small D};

    \draw (4.15, 6.75) -- (4.85, 6.50);
    \draw (4.15, 6.50) -- (4.85, 6.50);
    \draw (4.15, 6.00) -- (4.85, 6.75);

    \draw (5, 6.75) node {\small A};
    \draw (5, 6.50) node {\small B};

    \draw (5.15, 6.75) -- (5.85, 6.00);
    \draw (5.15, 6.50) -- (5.85, 6.25);

    \draw (6, 6.25) node {\small C};
    \draw (6, 6.00) node {\small D};

    \draw (6.15, 6.00) -- (6.85, 6.75);

    \draw (7, 6.70) node {\small A};

    % 110
    \draw (4.5, 3.5) node {\Symbol{1}};
    \draw (5.5, 3.5) node {\Symbol{1}};
    \draw (6.5, 3.5) node {\Symbol{0}};

    \draw (4, 4.75) node {\small A};
    \draw (4, 4.50) node {\small B};
    \draw (4, 4.00) node {\small D};

    \draw (4.15, 4.75) -- (4.85, 4.50);
    \draw (4.15, 4.50) -- (4.85, 4.50);
    \draw (4.15, 4.00) -- (4.85, 4.75);

    \draw (5, 4.75) node {\small A};
    \draw (5, 4.50) node {\small B};

    \draw (5.15, 4.75) -- (5.85, 4.50);
    \draw (5.15, 4.50) -- (5.85, 4.50);

    \draw (6, 4.50) node {\small B};

    \draw (6.15, 4.50) -- (6.85, 4.25);

    \draw (7, 4.25) node {\small C};

    % 111
    \draw (4.5, 1.5) node {\Symbol{1}};
    \draw (5.5, 1.5) node {\Symbol{1}};
    \draw (6.5, 1.5) node {\Symbol{1}};

    \draw (4, 2.75) node {\small A};
    \draw (4, 2.50) node {\small B};
    \draw (4, 2.00) node {\small D};

    \draw (4.15, 2.75) -- (4.85, 2.50);
    \draw (4.15, 2.50) -- (4.85, 2.50);
    \draw (4.15, 2.00) -- (4.85, 2.75);

    \draw (5, 2.75) node {\small A};
    \draw (5, 2.50) node {\small B};

    \draw (5.15, 2.75) -- (5.85, 2.50);
    \draw (5.15, 2.50) -- (5.85, 2.50);

    \draw (6, 2.50) node {\small B};

    \draw (6.15, 2.50) -- (6.85, 2.50);

    \draw (7, 2.50) node {\small B};
  \end{tikzpicture}
  \caption{All observable words of length 3 for the PSB Process. Each
    word has been annotated with the paths through which that word
    invokes synchrony. It is not until the observation of three
    symbols that in all cases there is only a single possible state.
    There are, however, some words which induce synchrony more
    quickly. }
\label{fig:sync_words}
\end{figure}
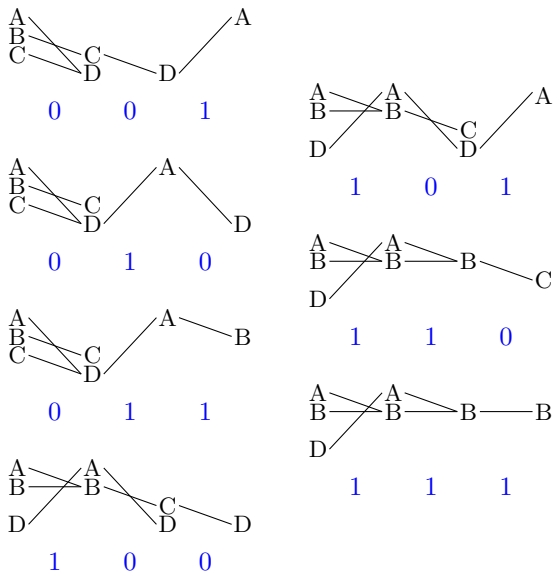

This provides an improved method of determining the Markov order.
Enumerate all words of increasing length noting which have
synchronized and which have not. When all the words at the current
length have synchronized, then that length is the Markov order
$\MOrder$. This procedure has been completed for the PSB Process in
Fig.~\ref{fig:sync_words}. It can be verified that at lengths $0$,
$1$, and $2$ it is possible to still have ambiguity as to which state
this system is in. For example, if the two symbols $10$ are observed,
the system may be in either state $C$ or state $D$. One more
observation is required to disambiguate which it is. Therefore, as
observed previously, the Markov order for this process is $R = 3$.
This method is improved by lexicographically enumerating words of
increasing length \emph{until} they synchronize to a single state. The
longest such word---a prefix-free synchronizing word---is the Markov
order $\MOrder$, since by that point every shorter word will have
synchronized and, therefore, the causal states will be determined
uniquely by words of that length.

This method addresses several weaknesses of the naive approach. Now,
neither $\EE$ nor $\hmu$ are needed, nor do we need to concern
ourselves with the details of comparing nearly equal numerical values.
However, the method relies on enumerating prefix-free synchronizing words,
and it is quite possible for a process to have an infinite  number of
prefix-free synchronizing words.  In these situations, it is not feasible to
enumerate them all, hoping to identify the longest. To address this problem,
we turn to formal language theory \cite{Hopcroft2001}.

\subsection{State Subset Construction}
\label{sec:subset-construction}

The remaining problem is to find the longest prefix-free synchronizing word
without having to enumerate them all. This can be accomplished with a standard
algorithm from the theory of finite automata. We construct an object known as
the \emph{power automaton} (PA), so-named since its states are elements of the
power set of a given automaton's states.

Construction of the power automaton begins with a single state: the
set of all states from the \eM. This is the PA's \emph{start}
state. Then recursively, for each state in the PA and each symbol,
consider all \eM\ states that can be reached by any \eM\ state
within the current PA state on the currently considered symbol.
A new PA state consisting of the set of \eM\ successor states is
added, along with a directed edge from the current to the new PA state,
labeled with the current symbol. Once the successors to each PA state
have been determined, there will be a subgraph of the PA that is
isomorphic to the recurrent \eM. This subgraph is the PA's
\emph{recurrent} component. When the \eM\ generates an ergodic
process, this subgraph is the only strongly connected component with
no outgoing edges. The remainder of the PA consists of \emph{transient}
states.

Synchronizing words are associated with particular PA paths.
Each path begins in the start state and traverses edges in PA's
transient portion. Eventually, the path continues to a PA
recurrent state. Prefix-free synchronizing words have paths that end as soon
as they reach a recurrent PA state. To find the longest prefix-free
synchronizing word, we weight each edge in PA's transient part
with the value $-1$ and each edge in the recurrent part with $0$.
With these modifications, the Bellman-Ford algorithm can be employed to
discover the path of least weight from the start state to any
recurrent state. Due to the weighting, the path of least weight is the
longest.

The alternative Floyd-Warshall algorithm can also be used;
see Ref.~\cite{Cormen2009} for details regarding both. We choose the
Bellman-Ford algorithm for two reasons. First, it works on graphs with
negative weight and, second, it detects negative-weight cycles. A
negative weight cycle here implies that the longest path is arbitrary
(infinite) in length.

This specifies a complete method for computing the Markov order
efficiently and accurately from a model of a process. First, construct
the power automaton. Then, weight the edges according to their status
as transient or recurrent. Last, find the path of least weight from
the start to a recurrent state. It runs in $O(\MeasAlphabet 2^{2N})$
time, which is exponential but finite. And, it depends only on integer
calculations. In this way, it circumvents all the computational
difficulties encountered in the naive approach. Thus, if one can infer
an accurate model from observations of a system, the problem of
computing that system's Markov order is solved.

This method also provides a solution to weakness (iv) of the naive
algorithm~(Sec.~\ref{sec:naive-approach}). When finite, the Markov order
depends on the longest path through the transient states of the power
automaton, and for an $n$ state recurrent \eM, there are at most
$f(n) \coloneqq 2^n - n - 1$
transient states (subtracting $n$ recurrent states and
also the empty set). Since loops in the transient structure imply
infinite Markov order, it follows that the longest possible path is one which
visits each of the transient states. Thus, if the Markov order has not been
found by $L = f(n)$, then it is safe to conclude that the Markov order
is infinite. Since, the Markov order bounds the cryptic order, the same
bound works for the cryptic order. It is an open problem to find a tight upper
bound for the Markov order in terms of the number of states and the number of
symbols in the alphabet.

\subsection{Examples}
\label{sec:examples}

\begin{figure}
  \centering
  \begin{tikzpicture}[style=vaucanson, bend angle=20]
    \node [state]      (A)                  {A};
    \node [state]      (B)    [right of=A]  {B};
    \node [state]      (C)    [below of=B]  {C};
    \node [state]      (D)    [below of=A]  {D};
    \node [state, red] (AB)   [above of=B]  {AB};
    \node [state, red] (CD)   [above of=A]  {CD};
    \node [state, red] (ABCD) [above of=CD] {ABCD};

    \path (ABCD) edge [red]                   node [swap] { \Symbol{0}} (CD)
          (ABCD) edge [red]                   node        { \Symbol{1}} (AB)
          (AB)   edge [red]                   node [swap] { \Symbol{0}} (CD)
          (AB)   edge [red]                   node [swap] { \Symbol{1}} (B)
          (CD)   edge [red, out=-135, in=135] node [swap] { \Symbol{0}} (D)
          (CD)   edge [red]                   node        { \Symbol{1}} (A)
          (A)    edge [bend left]             node        { \Symbol{1}} (B)
          (A)    edge [bend left]             node        { \Symbol{0}} (D)
          (B)    edge [loop above right]      node        {~\Symbol{1}} (B)
          (B)    edge [bend left]             node        { \Symbol{0}} (C)
          (C)    edge [bend left]             node        { \Symbol{0}} (D)
          (D)    edge [bend left]             node        { \Symbol{1}} (A);
  \end{tikzpicture}
\caption{PSB Process power automaton. The longest path beginning from state
  $ABCD$, traversing transient (red) edges, and ending in a recurrent (black)
  state is of length $3$: $ABCD \edge{1} AB \edge{0} CD \edge{1} A ( \text{or}
  \edge{0} D )$.
  }
\label{fig:mo1}
\end{figure}
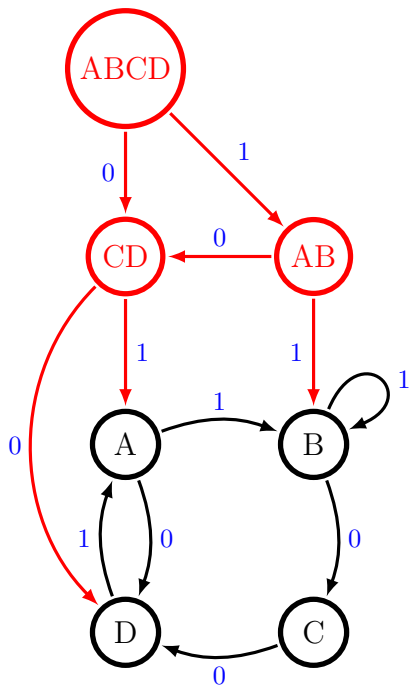

A variety of qualitatively different behaviors can be exhibited by the
Markov order algorithm. Here, we illustrate the typical cases.
Applying it to the PSB Process, the algorithm produces the fairly
simple transient structure consisting of three nodes---PA states
$ABCD$, $AB$, and $CD$---seen in Fig.~\ref{fig:mo1}. There are two
longest paths starting from PA start state $ABCD$ and ending in a
recurrent node: $ABCD \edge{1} AB \edge{0} CD \edge{1} A$, which is
traversed with the word $101$, and $ABCD \edge{1} AB \edge{0} CD
\edge{0} D$, traversed with the word $100$. This means that the
longest prefix-free synchronizing words are $101$ and $100$, both of
length three, and therefore PSB Process's Markov order is $R = 3$.

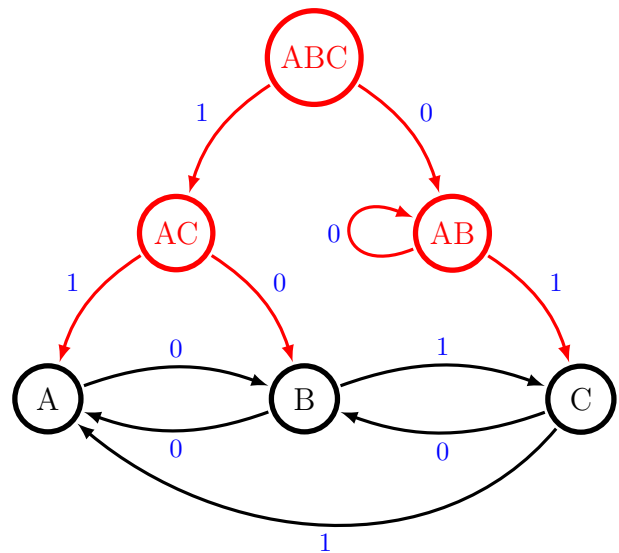
\begin{figure}
  \centering
  \begin{tikzpicture}[style=vaucanson, bend angle=20]
    \node [state, red] (ABC)                                    {ABC};
    \node [state, red] (AC)  [below left=1.5cm and 1cm of ABC]  {AC};
    \node [state, red] (AB)  [below right=1.5cm and 1cm of ABC] {AB};
    \node [state]      (A)   [below left=1.5cm and 1cm of AC]   {A};
    \node [state]      (B)   [below right=1.5cm and 1cm of AC]  {B};
    \node [state]      (C)   [below right=1.5cm and 1cm of AB]  {C};

    \path (ABC) edge [red, bend left]   node        {\Symbol{0}} (AB)
          (ABC) edge [red, bend right]  node [swap] {\Symbol{1}} (AC)
          (AC)  edge [red, bend left]   node        {\Symbol{0}} (B)
          (AC)  edge [red, bend right]  node [swap] {\Symbol{1}} (A)
          (AB)  edge [red, loop left]   node        {\Symbol{0}} (AB)
          (AB)  edge [red, bend left]   node        {\Symbol{1}} (C)
          (A)   edge [bend left]        node        {\Symbol{0}} (B)
          (B)   edge [bend left]        node        {\Symbol{0}} (A)
          (B)   edge [bend left]        node        {\Symbol{1}} (C)
          (C)   edge [bend left]        node        {\Symbol{0}} (B)
          (C)   edge [out=-130, in=-40] node        {\Symbol{1}} (A);
  \end{tikzpicture}
\caption{Typical complications in the PA for a finite-state non-Markovian
  Process.  The signature is the loop $AB \edge{0} AB$ in the transient
  structure. This means there is the possibility of an arbitrarily long series
  of observations that never synchronize and that, in turn, cause Markov order
  to diverge. Generically, loops in the transient structure can consist of more
  than one PA state.
  }
  \label{fig:mo2}
\end{figure}

The second process we analyze is shown in Fig.~\ref{fig:mo2}. It has a
slightly more complicated transient structure than that of the PSB
Process. Of particular note is the self-loop on PA state $AB$. This loop
exists because \eM\ states $A$ and $B$ transition to each other on
producing a $0$. As a consequence, we cannot determine the state until
observing a $1$. This inability to synchronize on some words results in a
non-Markovian process; that is, $\MOrder = \infty$. The Bellman-Ford algorithm
terminates as soon as it detects the corresponding negative-weight
cycle.

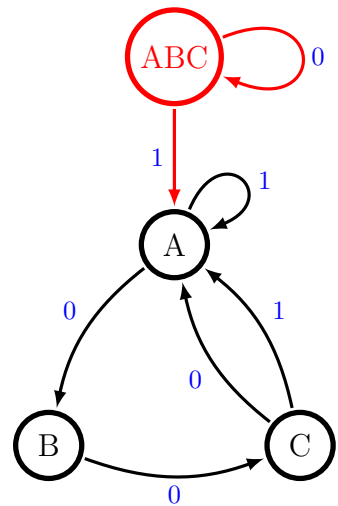
\begin{figure}
  \centering
  \begin{tikzpicture}[style=vaucanson, bend angle=20]
    \node [state, red] (ABC)                                {ABC};
    \node [state]      (A)   [below of=ABC]                 {A};
    \node [state]      (B)   [below left=2cm and 1cm of A]  {B};
    \node [state]      (C)   [below right=2cm and 1cm of A] {C};

    \path (ABC) edge [red, loop right]  node        { \Symbol{0}} (ABC)
          (ABC) edge [red]              node [swap] { \Symbol{1}} (A)
          (A)   edge [loop above right] node        {~\Symbol{1}} (A)
          (A)   edge [bend right]       node [swap] { \Symbol{0}} (B)
          (B)   edge [bend right]       node [swap] { \Symbol{0}} (C)
          (C)   edge [bend left]        node        { \Symbol{0}} (A)
          (C)   edge [bend right]       node [swap] { \Symbol{1}} (A);
  \end{tikzpicture}
  \caption{Like Fig.~\ref{fig:mo2}'s process, the Nemo Process here is
    non-Markovian. The Nemo Process make this perhaps clearer,
    however, since the recurrent states permute into each other upon
    observing a $0$. The transient structure makes this explicit:
    $ABC$ maps back to itself on a $0$. }
\label{fig:mo3}
\end{figure}

Our third example is the Nemo Process, shown in Fig.~\ref{fig:mo3}. Its
transient structure is particularly simple: a single state
representing all the recurrent states. Since the recurrent states
simply permute upon observing a $0$, the word $0000\ldots$ never
allows one to determine in which state the system is. This is
indicated by the self-loop on PA state $ABC$. This once again means
that the process is non-Markovian and has $\MOrder = \infty$. This
condition is detected by the algorithm as well.

\section{Cryptic Order}
\label{sec:cryptic-order}

We now turn to calculate a process's cryptic order $\COrder$. Recall that Eq.~(\ref{eq:CrypticOrder}) involves a condition on the \emph{infinite} future. With probability one, each infinite future synchronizes for exactly synchronizing \eMs~\cite{Travers2010a}. We can then consider the problem of calculating $\COrder$ to be that of determining as much of a state history as possible, given a prefix-free synchronizing word and the state to which it synchronized.  The maximum number of states we cannot retrodict is then the
cryptic order.

\subsection{Calculation}

\begin{figure}
  \centering
  \begin{tikzpicture}
    % \draw [help lines] (0, 0) grid (7, 8);

    % 001
    \draw (0.5, 6.5) node {\Symbol{0}};
    \draw (1.5, 6.5) node {\Symbol{0}};
    \draw (2.5, 6.5) node {\Symbol{1}};

    \draw (0, 7.50) node {\small B};

    \draw (0.15, 7.50) -- (0.85, 7.25);

    \draw (1, 7.25) node {\small C};

    \draw (1.15, 7.25) -- (1.85, 7.00);

    \draw (2, 7.00) node {\small D};

    \draw (2.15, 7.00) -- (2.85, 7.75);

    \draw (3, 7.75) node {\small A};

    % 010
    \draw (0.5, 4.5) node {\Symbol{0}};
    \draw (1.5, 4.5) node {\Symbol{1}};
    \draw (2.5, 4.5) node {\Symbol{0}};

    \draw (0, 5.75) node {\small A};
    \draw (0, 5.25) node {\small C};

    \draw (0.15, 5.75) -- (0.85, 5.00);
    \draw (0.15, 5.25) -- (0.85, 5.00);

    \draw (1, 5.00) node {\small D};

    \draw (1.15, 5.00) -- (1.85, 5.75);

    \draw (2, 5.75) node {\small A};

    \draw (2.15, 5.75) -- (2.85, 5.00);

    \draw (3, 5.00) node {\small D};

    % 011
    \draw (0.5, 2.5) node {\Symbol{0}};
    \draw (1.5, 2.5) node {\Symbol{1}};
    \draw (2.5, 2.5) node {\Symbol{1}};

    \draw (0, 3.75) node {\small A};
    \draw (0, 3.25) node {\small C};

    \draw (0.15, 3.75) -- (0.85, 3.00);
    \draw (0.15, 3.25) -- (0.85, 3.00);

    \draw (1, 3.00) node {\small D};

    \draw (1.15, 3.00) -- (1.85, 3.75);

    \draw (2, 3.75) node {\small A};

    \draw (2.15, 3.75) -- (2.85, 3.50);

    \draw (3, 3.50) node {\small B};

    % 100
    \draw (0.5, 0.5) node {\Symbol{1}};
    \draw (1.5, 0.5) node {\Symbol{0}};
    \draw (2.5, 0.5) node {\Symbol{0}};

    \draw (0, 1.75) node {\small A};
    \draw (0, 1.50) node {\small B};

    \draw (0.15, 1.75) -- (0.85, 1.50);
    \draw (0.15, 1.50) -- (0.85, 1.50);

    \draw (1, 1.50) node {\small B};

    \draw (1.15, 1.50) -- (1.85, 1.25);

    \draw (2, 1.25) node {\small C};

    \draw (2.15, 1.25) -- (2.85, 1.00);

    \draw (3, 1.00) node {\small D};

    % 101
    \draw (4.5, 5.5) node {\Symbol{1}};
    \draw (5.5, 5.5) node {\Symbol{0}};
    \draw (6.5, 5.5) node {\Symbol{1}};

    \draw (4, 6.00) node {\small D};

    \draw (4.15, 6.00) -- (4.85, 6.75);

    \draw (5, 6.75) node {\small A};

    \draw (5.15, 6.75) -- (5.85, 6.00);

    \draw (6, 6.00) node {\small D};

    \draw (6.15, 6.00) -- (6.85, 6.75);

    \draw (7, 6.70) node {\small A};

    % 110
    \draw (4.5, 3.5) node {\Symbol{1}};
    \draw (5.5, 3.5) node {\Symbol{1}};
    \draw (6.5, 3.5) node {\Symbol{0}};

    \draw (4, 4.75) node {\small A};
    \draw (4, 4.50) node {\small B};
    \draw (4, 4.00) node {\small D};

    \draw (4.15, 4.75) -- (4.85, 4.50);
    \draw (4.15, 4.50) -- (4.85, 4.50);
    \draw (4.15, 4.00) -- (4.85, 4.75);

    \draw (5, 4.75) node {\small A};
    \draw (5, 4.50) node {\small B};

    \draw (5.15, 4.75) -- (5.85, 4.50);
    \draw (5.15, 4.50) -- (5.85, 4.50);

    \draw (6, 4.50) node {\small B};

    \draw (6.15, 4.50) -- (6.85, 4.25);

    \draw (7, 4.25) node {\small C};

    % 111
    \draw (4.5, 1.5) node {\Symbol{1}};
    \draw (5.5, 1.5) node {\Symbol{1}};
    \draw (6.5, 1.5) node {\Symbol{1}};

    \draw (4, 2.75) node {\small A};
    \draw (4, 2.50) node {\small B};
    \draw (4, 2.00) node {\small D};

    \draw (4.15, 2.75) -- (4.85, 2.50);
    \draw (4.15, 2.50) -- (4.85, 2.50);
    \draw (4.15, 2.00) -- (4.85, 2.75);

    \draw (5, 2.75) node {\small A};
    \draw (5, 2.50) node {\small B};

    \draw (5.15, 2.75) -- (5.85, 2.50);
    \draw (5.15, 2.50) -- (5.85, 2.50);

    \draw (6, 2.50) node {\small B};

    \draw (6.15, 2.50) -- (6.85, 2.50);

    \draw (7, 2.50) node {\small B};
  \end{tikzpicture}
  \caption{Key paths for determining cryptic order $\COrder$: We start
    with the paths in Fig.~\ref{fig:sync_words}, except we remove
    paths that do not survive to the end of the sync word. The
    surviving paths give us the cryptic order: They each identify a
    single state by length $\ell = 2$ and so $\COrder = 2$. }
\label{fig:sync_words_cryptic}
\end{figure}
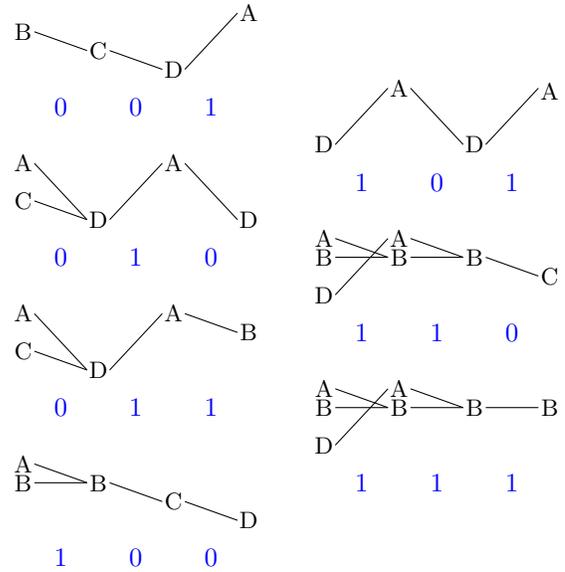

Figure~\ref{fig:sync_words_cryptic} depicts how the cryptic order is
determined. Only the paths in Fig.~\ref{fig:sync_words} that survive
all the way to synchrony (at the Markov order) are reproduced. From
these, we determine how many symbols into each word we must parse
(from the left) before the \eM\ is in one state only. The maximum such
length is the cryptic order $\COrder$.

As with the Markov order, we need only consider prefix-free
synchronizing words. However, we are again faced with the prospect
that there may be an infinite number of prefix-free synchronizing
words. Fortunately, a better method is available, and it too begins by
constructing the power automaton. Now, we examine the ``veracity'' of
each transient edge. Take as an example the edge $ABC \edge{1} A$ in
Fig.~\ref{fig:mo3}. It states that upon producing a $1$ from the
superposition of states $A$, $B$, and $C$, the system can only
transition to state $A$. For the cryptic order, we now condition on
the fact that we are in state $A$ and ask what states could have
transitioned to $A$ on a $1$. Upon inspection, its clear that the
system could have only transitioned from states $A$ or $C$ on a $1$.
The core of the cryptic order algorithm is to inspect each transient
edge in the power automaton in this manner, updating the PA's
structure to ``honestly'' reflect the process's dynamics. In this
instance, we create a state $AC$ that transitions to state $A$ on a
$1$ instead of transitioning from $ABC$ on a $1$.

After creating a state, the automaton must be made consistent. To do this,
subset construction is applied to include any newly added states.  Generally,
this creates new edges as well. And, these too must be analyzed by the cryptic
order algorithm. Once every edge has been inspected, some transient structure
will remain. Once again, the longest path is the key, and the same
edge-weighting method (Bellman-Ford) is employed to find it and so give the
cryptic order.

\subsection{Examples}
\label{sec:examples-1}

The ways in which the cryptic order algorithm modifies the power
automaton are diverse. Each example from Sec.~\ref{sec:examples} above
illustrates a different behavior.

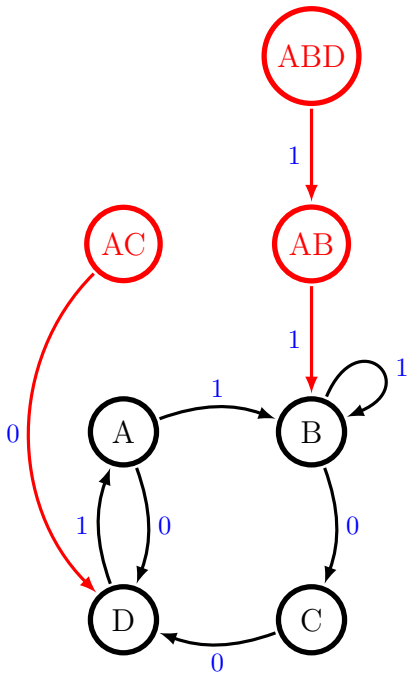
\begin{figure}
  \centering
  \begin{tikzpicture}[style=vaucanson, bend angle=20]
    \node [state]      (A)                 {A};
    \node [state]      (B)   [right of=A]  {B};
    \node [state]      (C)   [below of=B]  {C};
    \node [state]      (D)   [below of=A]  {D};
    \node [state, red] (AB)  [above of=B]  {AB};
    \node [state, red] (AC)  [above of=A]  {AC};
    \node [state, red] (ABD) [above of=AB] {ABD};

    \path (ABD) edge [red]                   node [swap] { \Symbol{1}} (AB)
          (AB)  edge [red]                   node [swap] { \Symbol{1}} (B)
          (AC)  edge [red, out=-135, in=135] node [swap] { \Symbol{0}} (D)
          (A)   edge [bend left]             node        { \Symbol{1}} (B)
          (A)   edge [bend left]             node        { \Symbol{0}} (D)
          (B)   edge [loop above right]      node        {~\Symbol{1}} (B)
          (B)   edge [bend left]             node        { \Symbol{0}} (C)
          (C)   edge [bend left]             node        { \Symbol{0}} (D)
          (D)   edge [bend left]             node        { \Symbol{1}} (A);
  \end{tikzpicture}
\caption{Cryptic order algorithm applied to the PSB Process: The power
  automaton in Fig.~\ref{fig:mo1} suggests that the word $11$ could originate
  in any of $A$, $B$, $C$, or $D$. Careful inspection of the recurrent
  structure, though, shows that $C$ cannot be the originator of $11$, whereas
  the other three states can. The cryptic order algorithm accounts for such
  constraints.  The longest path from a transient state to a recurrent state is
  $ABD \edge{1} AB \edge{1} B$ and, therefore, $\COrder = 2$.
  }
\label{fig:co1}
\end{figure}

First, consider its behavior on the PSB Process (Fig. \ref{fig:mo1}), the
final result of which is shown in Fig.~\ref{fig:co1}. The edge $CD \edge{0} D$
in Fig. \ref{fig:mo1} can be
removed since it does not represent a path that is true. To see why,
note that to get to $D$ on a $0$, one must come from either state $A$
or state $C$. However, since we are assuming $CD$, the process must be
in either state $C$ or $D$. The intersection of those two sets is
state $C$ and it is, therefore, the only possible state the system
could have \emph{actually} been in. Thus, $CD \edge{0} D$ is a
misrepresentation from the cryptic order perspective and, in fact,
it corresponds to the edge $C \edge{0} D$, which already exists in the
PA. So, the edge $CD \edge{0} D$ is removed.

This is not all, however.
We must maintain the path's provenance. The edges that came in to $CD$
must be redirected to $C$ (add edges $ABCD \edge{0} C$ and $AB
\edge{0} C$), since those are the edges that would have been traversed
immediately prior to $CD \edge{0} D$. Note that these edges are later
removed in this recursive algorithm and so do not appear in
Fig.~\ref{fig:co1}. In the end, we see that the longest path from a
start state to the recurrent states is $2$ and, therefore, $\COrder =
2$, one less than the Markov order $\MOrder = 3$.

\begin{figure}
  \centering
  \begin{tikzpicture}[style=vaucanson, bend angle=20]
    \node [red]        (oABC)                       {};
    \node [red]        (oAC)  [below left of=oABC]  {};
    \node [red]        (oAB)  [below right of=oABC] {};
    \node [state]      (A)    [below left of=oAC]   {A};
    \node [state]      (B)    [below right of=oAC]  {B};
    \node [state]      (C)    [below right of=oAB]  {C};
    \node [state, red] (AC)   [above of=B]          {AC};
    \node [state, red] (BC)   [above of=AC]         {BC};

    \path (BC)  edge [red]              node {\Symbol{1}} (AC)
          (AC)  edge [red]              node {\Symbol{0}} (B)
          (A)   edge [bend left]        node {\Symbol{0}} (B)
          (B)   edge [bend left]        node {\Symbol{0}} (A)
          (B)   edge [bend left]        node {\Symbol{1}} (C)
          (C)   edge [bend left]        node {\Symbol{0}} (B)
          (C)   edge [out=-130, in=-40] node {\Symbol{1}} (A);
  \end{tikzpicture}
  \caption{Cryptic order analysis of Fig.~\ref{fig:mo2}'s process: The
    transient structure branch shown there---$ABC \edge{0} (AB
    \edge{0} AB)^* \edge{1} C$, with the arbitrarily long
    synchronizing word $00^*1$---can be perfectly retrodicted.
    Moreover, only a fragment of the left branch of the transient
    structure remains. This fragment has a length of $2$, and so
    $\COrder = 2$. }
\label{fig:co2}
\end{figure}
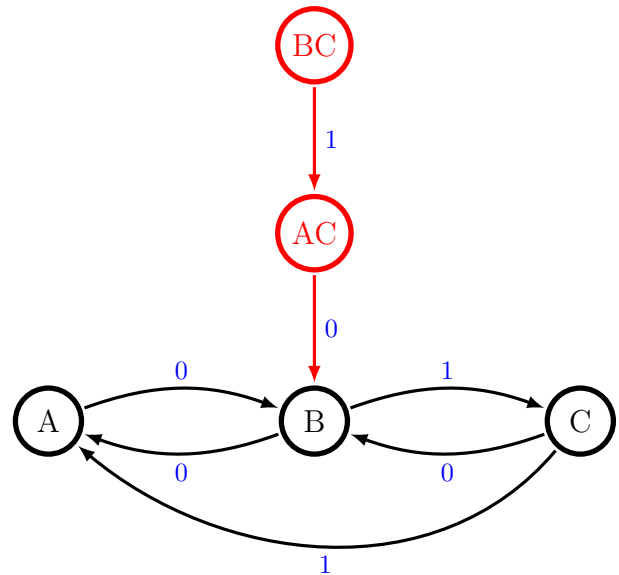

Next, consider the example from Fig.~\ref{fig:mo2}. The final output of the
cryptic order algorithm is shown in Fig.~\ref{fig:co2}. This process's
PA consists of two major branches: One with a maximum depth of $2$ and
the other containing a loop. The cryptic order algorithm discovers
that the branch with a loop is completely retrodictable. $AB \edge{1}
C$ is actually $B \edge{1} C$, and this creates edges $AB \edge{0} B$
and $ABC \edge{0} B$, again to maintain provenance. The first of these
newly added edges is also retrodictable: $AB \edge{0} B$ can only be
$A \edge{0} B$. The second, $ABC \edge{0} B$, is in fact $AC \edge{0}
B$. Along this branch of the transient structure, we are thus only
unable to retrodict the word $01$, of which the $1$ can be
retrodicted, simply leaving us with $AC \edge{0} B$. The previous
branch is more easily analyzed, leaving us with $BC \edge{1} AC
\edge{0} B$, the later part of which was already in the PA from
analyzing the other branch. This leaves a longest path of length $2$,
making $\COrder = 2$. Thus, we see that this process is an example
with infinite Markov order, but finite cryptic order.

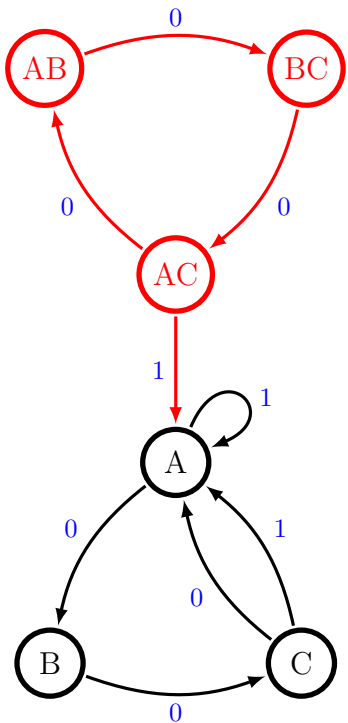
\begin{figure}
  \centering
  \begin{tikzpicture}[style=vaucanson, bend angle=20]
    \node [state, red] (AC)                                 {AC};
    \node [state, red] (AB) [above left=2cm and 1cm of AC]  {AB};
    \node [state, red] (BC) [above right=2cm and 1cm of AC] {BC};
    \node [state]      (A)  [below of=AC]                   {A};
    \node [state]      (B)  [below left=2cm and 1cm of A]   {B};
    \node [state]      (C)  [below right=2cm and 1cm of A]  {C};

    \path (AC) edge [red, bend left]   node        { \Symbol{0}} (AB)
          (AC) edge [red]              node [swap] { \Symbol{1}} (A)
          (AB) edge [red, bend left]   node        { \Symbol{0}} (BC)
          (BC) edge [red, bend left]   node        { \Symbol{0}} (AC)
          (A)  edge [loop above right] node        {~\Symbol{1}} (A)
          (A)  edge [bend right]       node [swap] { \Symbol{0}} (B)
          (B)  edge [bend right]       node [swap] { \Symbol{0}} (C)
          (C)  edge [bend left]        node        { \Symbol{0}} (A)
          (C)  edge [bend right]       node [swap] { \Symbol{1}} (A);
  \end{tikzpicture}
\caption{Cryptic order analysis of the Nemo Process: Its power automaton
  (Fig. \ref{fig:mo3}) contains the edge $ABC \edge{1} A$. However, upon closer
  inspection only states $A$ and $C$ can transition to $A$ on a $1$.  This
  creates the $AC$ state. When emitting a $0$, $AC$ becomes $AB$ and on a
  second $0$ that becomes $BC$. A third $0$ completes the cycle. The edges
  indicate legitimate transitions as well: States that actually lead to $AC$ on
  a $0$ are $BC$ and those that lead to $BC$ are $AB$, and so on. This leads to
  a cycle in the cryptic order algorithm's calculated transient structure.
  Therefore, one concludes that $\COrder = \infty$.
  }
\label{fig:co3}
\end{figure}

The last example to consider is the Nemo Process. Recall that it is infinite
Markov, as observed in Fig.~\ref{fig:mo3}. Applying the cryptic order algorithm
results in the structure shown in Fig.~\ref{fig:co3}. In this case, the
transient structure grows under the algorithm. The edge $ABC \edge{1} A$,
connecting the transient to the recurrent structure in the power automaton, is
modified by the algorithm since $B$ cannot transition to $A$ on a $1$. The
state $AC$ is created and connected to $A$. Completing the power automaton
structure from this state results in states $AB$ and $BC$ being added, forming
the cycle $AC \edge{0} AB \edge{0} BC \edge{0} AC$. The algorithm terminates
when the cycle is detected in this way. The cycle is valid as far as the
cryptic order is concerned: Each of its states can be transitioned to from the
recurrent state associated with the prior state in the cycle. The cycle results
in an arbitrarily long path and, therefore, $\COrder = \infty$.

\section{Other Natural Time Scales}
\label{sec:other-orders}

Paralleling the interpretation of the Markov and cryptic orders as the
block lengths at which an associated information measure reaches its
asymptotic behavior, this section briefly defines several new time scales
associated with the multivariate information measures recently
introduced in Ref.~\cite{James2011} to dissect the information in a
single measurement.

The first order $k_I$ is the length at which the multivariate mutual
information $I[\MeasSymbol_0; \MeasSymbol_1; \ldots ; \MeasSymbol_{N-1}]$
reaches its asymptotic behavior. Unfortunately, no bounds
are known for this order.

The next collection of time scales---denoted $k_R$, $k_B$, $k_Q$, and $k_W$---are
the lengths at which the residual entropy $\rmu$, bound information
$\bmu$, enigmatic information, and local exogenous information each
reach their respective asymptotes~\cite{James2011}. Furthermore, these
four orders are equal, due to the linear interdependence of their
respective measures. It turns out that there are lower and upper
bounds for these with respect to the Markov order, which can be easily
explained. Consider Fig.~8 in Ref.~\cite{James2011}: By definition
$\H[\MeasSymbols{}{0}]$ can be replaced with
$\H[\MeasSymbols{-\MOrder}{0}]$ and, if the process is stationary,
$\H[\MeasSymbols{1}{}]$ with $\H[\MeasSymbols{1}{\MOrder+1}]$. It is
therefore reasonable that one requires at least $\MOrder$ symbols and
most $2\MOrder$ symbols to accurately dissect $\H[\MeasSymbol_0]$. In
fact, numerical surveys that we have carried out agree with these
limits.

Finally, a sequel analyzes the elusive information $\sigmamu$, showing that
the Markov order $R$ equals the length $k_{\sigmamu}$ at which the present
measurement block $\MeasSymbols{0}{\ell}$ renders the past and future
conditionally independent.

While we have defined these orders and provided bounds, it remains to
be seen if there exist efficient methods to calculate them, let alone
topological interpretations for each.

\section{Survey}
\label{sec:survey}

We illustrate the above results and algorithms, and their usefulness,
by empirically answering several simple, but compelling questions
about the space of finitary processes. In particular, how typical are
infinite Markov order and infinite cryptic order?

\begin{figure}
  \centering
  \includegraphics[width=\columnwidth]{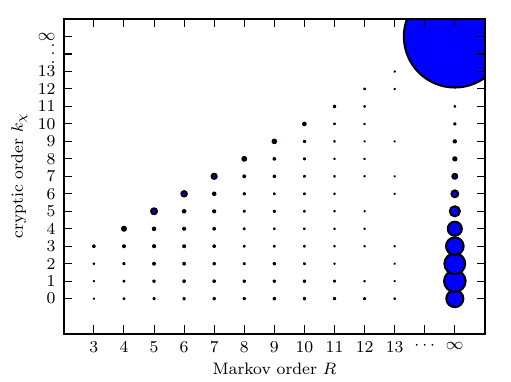}
  \caption{Distribution of Markov order $\MOrder$ and cryptic order
    $\COrder$ for all $1,132,613$ six-state, binary-alphabet,
    exactly-synchronizing \eMs. Marker size is proportional to the
    number of \eMs\ within this class at the same ($\MOrder$,
    $\COrder$). }
\label{fig:rvsk}
\end{figure}

Restricting ourselves to topological \eMs---those \eMs\ with a
distinct set of allowed transitions and equiprobable transition
probabilities---we enumerate all binary-alphabet processes with a
given number of states to which one can exactly synchronize. Ref.
\cite{Johnson2010} details their definition, the enumeration
algorithm, and how it gives a view of the space of structured
stochastic processes. For each of these \eMs, we compute its Markov
and cryptic orders. The result for all of the $1,132,613$ six-state
\eMs\ is shown in Fig.~\ref{fig:rvsk}.

The number of \eMs\ that share a ($\MOrder$, $\COrder$) pair is
encoded by the size of the circle at that ($\MOrder$, $\COrder$). The
vast majority of processes---in fact, $98\%$---are non-Markovian at
this state-size ($6$ states). Furthermore, most ($85\%$) of those
non-Markovian processes are also $\infty$-cryptic. However, this does
not imply that synchronization is difficult; quite the contrary:
synchronization occurs exponentially quickly~\cite{Trav10a}. What this
does mean is that with growing state size it becomes predominately
likely that a given process has particular sequences which will not
induce state synchronization.

Also of interest are the ``forbidden'' ($\MOrder$, $\COrder$) pairs
within the space of $6$-state topological \eMs. For example, \eMs\
with $\COrder = 4, 5, 8, 10, 11$ do not occur with $\MOrder = 13$.
Also, processes with infinite Markov order and finite cryptic order
appear to have a maximum cryptic order of $\COrder = 11$, despite the
fact that larger finite cryptic orders exist for finite Markov-order
processes.

\section{Spin Chains and Beyond}
\label{sec:spin-systems}

Although our primary goal was to precisely define length scales, several
being new, and to present efficient calculation methods for them, it will be
helpful to briefly draw out the physical meaning of Markov and cryptic orders
by analyzing their role in spin chains and related systems. (A sequel will
delve into this topic in greater depth.)

To start, recall that Ref.  ~\cite{Feldman1998} showed that the Markov order
$\MOrder$ of an \eM\ representing a (one-dimensional) Ising spin system is
upper bounded by the interaction range specified in a system's Hamiltonian.
Consider first the ferromagnetic, one-dimensional, nearest-neighbor Ising model
at different temperatures $T$. The \eMs\ for this family of systems are shown
in Fig.~\ref{fig:ferromagnetic}. As just noted, since the system has
nearest-neighbor interactions, the Markov order should be $\MOrder = 1$. This
is straightforward to see from the first \eM, which is for a finite temperature
$T$. Without an observation there are two possible causal states the system
could be in: $\uparrow$ or $\downarrow$. Once a single spin has been observed,
however, the causal state is known exactly.  This changes markedly at the
temperature limits, though. At $T = 0$, the system is in a ground configuration
of either all up spins or all down spins. Without an external field to break
this symmetry an observation must be made to determine in which of these states
it is and so the Markov order is still $\MOrder = 1$. In the presence of an
external field, however, there is only a single ground state---that aligned
with the field---and no observation is required to know in which state the
system is. Thus, $\MOrder = 0$. At $T = \infty$ the system collapses to a
single causal state where the next spin is entirely determined by thermal
fluctuations and so the Markov order is $\MOrder = 0$.

As a second case, consider the antiferromagnetic, one-dimensional,
nearest-neighbor Ising model, which is similar enough that it makes for a
useful contrast; see Fig.~\ref{fig:antiferromagnetic}. The finite temperature
and high temperature limits are identical to the ferromagnetic case, but the
low temperature case differs. At $T = 0$ the spin system forms a perfect
crystal of alternating spins and so one must make a single observation to know
in which spatial-phase the crystal is. Then, the entire structure is known
exactly.  Thus, the Markov order is $\MOrder = 1$. This situation is not a
broken symmetry as in ferromagnetic low-temperature case. Even with a nonzero
external field, an observation is still required to know in which causal state
the system is.

Overall, now that we can directly determine intrinsic lengths in
configurations, we see that the coupling range specified by a Hamiltonian need
not be an intrinsic property of realized configurations. The simple extremes
above make this easy to understand. At infinite temperature each system
configuration is equally likely: the Hamiltonian range has no effect on which
configurations are realized. At zero temperature only the ground states are
expressed and these need not explore all the possible configurations allowed by
the Hamiltonian.  Both of these situations mask the coupling range specified by
the Hamiltonian. Due to this, the Markov order $\MOrder$ captures the effective
coupling range and need not match that specified by the Hamiltonian.

In Ising spin chains, the cryptic order equals the Markov order.
(This is due most directly to the fact that spin blocks are in one-to-one
correspondence with the \eM\ causal states. In addition, one must
add the caveat that the \eM\ be ergodic.)
This equality need not be the case, however, even in simple physical systems.
We note how restrictive Hamiltonian-specified dynamics are via two
(again, 1D) examples of infinite Markov order, but finite cryptic
order, that arise from finite specification. In the
first class of systems, even though one starts with strictly local
interactions---configurations with finite Markov order specified by a
Hamiltonian with finite coupling range---a 1D system can anneal to one
with effectively infinite-range interactions, as shown in Ref. \cite{Varn03c}.
(See the \eM\ in Fig. 2 there.) In this particular case, the annealed
state is non-Markovian, exhibiting infinite-range structure and Markov
order $\MOrder = \infty$. Notably, the annealed configurations for
this example have finite cryptic order $\COrder = 4$. For the second
class of systems we just briefly note that these unusual length-scale
properties are not restricted to classical systems. They also arise in
quantum systems. See the analyses in Refs. \cite{Wies06c} and
\cite{Wies07a}.

Finally, since the results here emphasize properties intrinsic to realized
configurations, let's turn the question around. Given a single typical instance
from the ensemble of allowed configurations, how much can be inferred about the
Hamiltonian? Though the topological techniques described above do not provide
coupling amplitudes and the like, they do give the maximum range of effective
interactions. What does one do, though, without a Hamiltonian or some other
system specification? It turns out that a variety of methods exist for
inferring hidden Markov models from a sample. And, since any hidden Markov
model can be converted to an \eM\ \cite{Ellison2009}, from there the Markov and
cryptic orders can be directly computed. And so, the above methods can be
applied to a wide range of theoretically modeled or experimentally realized
physical systems.

\newcommand{\prob}{\frac{1}{2}\left(1 + \tanh{\beta}\right)}

\begin{figure}
  \centering
  \begin{tikzpicture}[style=vaucanson, bend angle=20]
    \draw (1.25, 1.25) node {$0 < T < \infty$};
    \draw (0, -1.25) node {};

    \node [state] (up)                 {$\uparrow$};
    \node [state] (down) [right of=up] {$\downarrow$};

    \path (up)   edge [loop left]  node {\Edge{p}{\uparrow}}     (up)
          (up)   edge [bend left]  node {\Edge{1-p}{\downarrow}} (down)
          (down) edge [bend left]  node {\Edge{1-p}{\uparrow}}   (up)
          (down) edge [loop right] node {\Edge{p}{\downarrow}}   (down);

  \end{tikzpicture}\\
  \begin{tikzpicture}[style=vaucanson, bend angle=20]
    \draw (1.25, 1.25) node {$T = 0$};
    \draw (0, -1.25) node {};

    \node [state] (up)                 {$\uparrow$};
    \node [state] (down) [right of=up] {$\downarrow$};

    \path (up)   edge [loop left]  node {\Edge{1}{\uparrow}}   (up)
          (down) edge [loop right] node {\Edge{1}{\downarrow}} (down);
  \end{tikzpicture}\\
  \begin{tikzpicture}[style=vaucanson, bend angle=20]
    \draw (0, 1.25) node {$T = \infty$};
    \draw (0, -1.25) node {};

    \node [state] (updown) {$\uparrow \downarrow$};

    \path (updown) edge [loop left]  node {\Edge{\half}{\uparrow}}   (updown)
          (updown) edge [loop right] node {\Edge{\half}{\downarrow}} (updown);
  \end{tikzpicture}
  \caption{\EMs\ for a one-dimensional ferromagnetic Ising model as a
    function of temperature $T$, where $p = \prob$, the external field
    $B = 0$, and $J = k_B = 1$. }
\label{fig:ferromagnetic}
\end{figure}
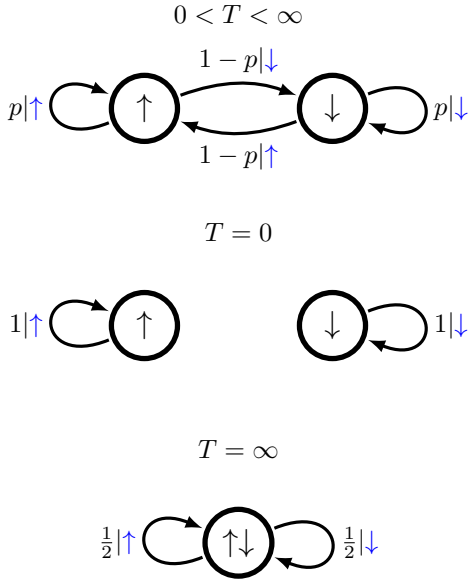

\begin{figure}
  \centering
  \begin{tikzpicture}[style=vaucanson, bend angle=20]
    \draw (1.25, 1.25) node {$0 < T < \infty$};
    \draw (0, -1.25) node {};

    \node [state] (up)                 {$\uparrow$};
    \node [state] (down) [right of=up] {$\downarrow$};

    \path (up)   edge [loop left]  node {\Edge{1-p}{\uparrow}}   (up)
          (up)   edge [bend left]  node {\Edge{p}{\downarrow}}   (down)
          (down) edge [bend left]  node {\Edge{p}{\uparrow}}     (up)
          (down) edge [loop right] node {\Edge{1-p}{\downarrow}} (down);

  \end{tikzpicture}\\
  \begin{tikzpicture}[style=vaucanson, bend angle=20]
    \draw (1.25, 1.25) node {$T = 0$};
    \draw (0, -1.25) node {};

    \node [state] (up)                 {$\uparrow$};
    \node [state] (down) [right of=up] {$\downarrow$};

    \path (up)   edge [bend left]  node {\Edge{1}{\downarrow}} (down)
          (down) edge [bend left]  node {\Edge{1}{\uparrow}}   (up);
  \end{tikzpicture}\\
  \begin{tikzpicture}[style=vaucanson, bend angle=20]
    \draw (0, 1.25) node {$T = \infty$};
    \draw (0, -1.25) node {};

    \node [state] (updown) {$\uparrow \downarrow$};

    \path (updown) edge [loop left]  node {\Edge{\half}{\uparrow}}   (updown)
          (updown) edge [loop right] node {\Edge{\half}{\downarrow}} (updown);
  \end{tikzpicture}
  \caption{\EMs\ for a one-dimensional antiferromagnetic Ising model
    as a function of temperature $T$. Here, $p = \prob$, the external
    field $B = 0$, and $J = k_B = 1$. }
\label{fig:antiferromagnetic}
\end{figure}
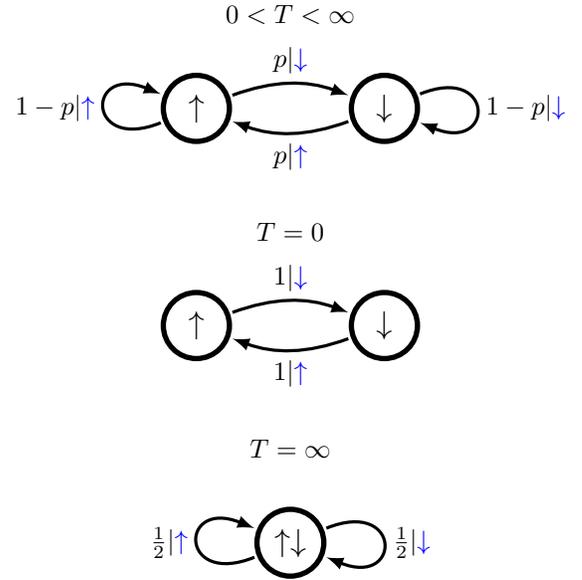

\section{Conclusion}
\label{sec:conclusion}

We began by defining two different measures of memory in complex systems. The
first, the Markov order $\MOrder$, is the length of time one must observe a
system in order to make accurate predictions of its behavior. The second, the
cryptic order $\COrder$, quantifies the ability to retrodict a system's
internal dynamics. We showed that despite their statistical nature, these time
scales are topological properties---properties of the synchronizing words of a
process's \eM.

We demonstrated how to compute these length scales for hidden Markov models,
most of which can be motivated in terms of the synchronization properties of
the underlying process. Interestingly, we found that one of
the most fundamental and important properties---the Markov order $\MOrder$---is
computable using \emph{only} the process's \eM. When calculated with non-\eMs,
the algorithms yield related quantities, such as the synchronization order. For
more details, see the Appendices. In addition, the \eM\ provides an exact
method for computing the cryptic order. From these results, we constructed very
efficient algorithms for their calculation.

In the empirical setting, we now see that one should first infer the \eM\ and
then, from it, calculate the Markov and cryptic orders. There are a number of
methods of inferring an \eM\ from data; e.g.,
Ref.~\cite[and citations therein]{Stre13a}. In the theoretical setting,
given some formal description of a process---such as a Hamiltonian or
general hidden Markov model---one can analytically calculate a process's \eM.
In any case, as soon as one has the \eM\, the preceding gives exact results.

To appreciate what is typical about these length scales, we surveyed the range
of Markov and cryptic orders in the space of all structured binary processes
represented by \eMs\ with six states. The main result was rather surprising,
infinite Markov and cryptic orders dominate. Thus, the topological analysis
leads one to conclude that synchronization, even to finite-state stochastic
processes, is generically difficult. However, from a probabilistic view it is
exponentially fast \cite{Trav10a,Trav10b}. A way to resolve this seeming
contradiction is to conjecture that the topological properties are driven by
sequences whose relative proportion vanishes with increasing length.  The
survey also revealed a variety of interesting ancillary properties that pose a
number of open questions, presumably combinatoric and group-theoretic in
nature.

We closed analyzing the role these scales play in classical (and briefly
quantum) spin systems, drawing out the physical interpretations. We emphasized,
in particular, the difference between the interaction range specified by a
Hamiltonian and the effective range of correlation in realized spin
configurations. This led us to propose calculating the orders to put
constraints on spin systems whose Hamiltonians are unknown.

Finally, appendices prove key claims above, discuss other related measures of
synchronization, survey the synchronization time and synchronization entropy,
and provide step-by-step details for each algorithm.

\section*{Acknowledgments}

We thank Ben Johnson and Nick Travers for many invaluable discussions.  This
work was partially supported by ARO grants W911NF-12-1-0234 and by
W911NF-12-1-0288 and by the Defense Advanced Research Projects Agency (DARPA)
Physical Intelligence project. The views, opinions, and findings contained here
are those of the authors and should not be interpreted as representing the
official views or policies, either expressed or implied, of ARO, DARPA, or the
Department of Defense.

\bibliography{orders}

\appendix

\section{Definitions}

Here, we provide additional results on length scales and synchronization and
prove a number of claims made in the main text.  First, we lay out the
definitions needed and then give several key results that follow. Building on
these, we delineate the central algorithms and conclude with a comparison of
synchronization time and synchronization entropy, notions perhaps more familiar
than other measures used and their length scales.

\subsection{Minimal Synchronizing Words}
\label{sec:msw}

For the synchronization problem, we consider an observer who begins
with a correct model (a presentation) of a process. The observer,
however, has no knowledge of the process's internal state. The
challenge is to analyze how an observer's knowledge of the internal
state changes as more and more measurements are observed.

At first glance, one might say that the observer's knowledge should
never decrease with additional measurements, corresponding to a
never-increasing state uncertainty, but this is generically not true.
In fact, it is possible for the observer's knowledge (measured in
bits) to oscillate with each new measurement. The crux of the issue is
that additional measurements are being used to inquire about the
\emph{current} state rather than the state at some fixed moment in
time.

It is helpful to identify the set of words that take the observer from
the condition of total ignorance to exactly knowing a process's state.
First, we introduce what we mean by synchronization in terms of lack
of state uncertainty. Second, we define the set of minimal
synchronizing words.

\begin{Def}
  A word $w$ of length $L$ is \emph{synchronizing} if the Shannon
  entropy over the internal state, conditioned on $w$, is zero:
\begin{align}
   Sync(w)
   \quad \Leftrightarrow \quad
   H[\CausalState_\ell | \MeasSymbols{0}{\ell}=w] = 0 ~,
\end{align}
where $Sync(w)$ is Boolean function.
\label{def:sw}
\end{Def}

\begin{Def}
  A presentation's set of \emph{minimal synchronizing words} is the
  set of synchronizing words that have no synchronizing prefix:
\begin{align*}
\Lsync \equiv \{ w ~ & | ~ Sync(w) \mathrm{~and~}
  \neg Sync(u) \mathrm{~for~all~} u: w = uv \}
  .
\end{align*}
\label{def:msw}
\end{Def}

\begin{Rem}
  $\Lsync$ is a prefix-free, regular language. If each word is
  associated with its probability of being observed, we obtain a
  prefix-free code encoding each path to synchrony---a word in
  $\Lsync$---with the associated probability of synchronizing via that
  path. These codes are generally nonoptimal in the familiar
  information-theoretic sense.
\end{Rem}

\subsection{Synchronization Order}

According to Sec.~\ref{sec:msw}, one is synchronized to a process's
presentation after seeing word $w$ if there is complete certainty in
the state. We now expand this view slightly to ask about
synchronization over all words of a particular length. Equivalently,
we examine synchronization to an ensemble of process realizations.
\begin{Def}
  The \emph{synchronization order} $\SOrder$ \cite{Crutchfield2010} is
  the minimum length for which every allowed word is a synchronizing
  word:
\begin{align}
  \SOrder \equiv \min \{ \ell ~|~
  H[ \CausalState_\ell | \MeasSymbols{0}{\ell} ] = 0 \} ~.
\end{align}
As for the Markov and cryptic orders, $\SOrder$ is considered $\infty$
when the condition does not hold for any finite $\ell$.
\label{def:so}
\end{Def}

\section{Results}
\label{sec:thes}

We now provide several results related to these length scales that
shed light on their nature, introducing connections and simplifications
that make their computation tractable.

\begin{Prop}
The synchronization order is:
\begin{align}
  \SOrder = \ max \{ \MOrder , \COrder \} ~.
\end{align}
\label{the:soemax}
\end{Prop}

\begin{ProProp}
First, note that:
\begin{align}
  H[ \CausalState_\ell | \MeasSymbols{0}{\ell} ] =
  H[ \MeasSymbols{0}{\ell} , \CausalState_\ell ] - H[ \MeasSymbols{0}{\ell} ]
  ~.
\end{align}
Since the block-state entropy upper bounds the block entropy, the
conditional entropy above can only reach its asymptotic value once
both terms have individually reached their asymptotic behavior. The
latter are controlled by $\COrder$ and $\MOrder$, respectively. \qed
\label{pro:soemax}
\end{ProProp}

This result reduces the apparent diversity of length scales,
eventually allowing one to calculate the Markov order via the
synchronization order, which itself is directly computable.

\begin{Prop}
For \eMs:
\begin{align}
  \MOrder = \SOrder ~.
\end{align}
\label{the:moeso}
\end{Prop}

\begin{ProProp}
  Applying the causal equivalence relation $\CausalEquivalence$ to
  Def.~\ref{eq:MarkovOrder} we find:
\begin{align}
  \Prob( \Future | \Past ) & = \Prob( \Future | \MeasSymbols{-\ell}{0}
  ) \implies \Past \,\CausalEquivalence\, \MeasSymbols{-\MOrder}{0} ~.
\end{align}
This further implies that the causal states $\CausalStateSet$ are
completely determined by $\MeasSymbols{-\MOrder}{0}$:
\begin{align}
  H[ \CausalState_0 | \MeasSymbols{-\MOrder}{0} ] = 0 ~.
\end{align}
This statement is equivalent to the Markov criterion. \qed
\label{pro:moeso}
\end{ProProp}

\begin{Rem}
  This provides an alternate proof that the cryptic order $\COrder$ is
  bounded above by the Markov order $\MOrder$ in an \eM\ via a simple
  shift in indices:
\begin{align}
  & H[ \CausalState_0 | \MeasSymbols{-\MOrder}{0} ] = 0 \\
  \implies & H[ \CausalState_\MOrder | \MeasSymbols{0}{\MOrder}] = 0 \\
  \implies & H[ \CausalState_\MOrder | \Future ] = 0 ~.
\end{align}
\end{Rem}

This proposition gives indirect access to the Markov order via a
particular presentation---the \eM. Since the Markov order is not
defined as a property of a presentation it would generally be
unobtainable, but due to unique properties of the \eM, it can be
accessed through the synchronization order.

There is a subclass of \eMs\ to which one synchronizes in finite time;
these are the \emph{exact} \eMs\ of Ref. \cite{Travers2010a}.

\begin{Prop}
  Given an exact \eM\ with finite Markov order $\MOrder$, the subshift
  of finite type that underlies it has a ``step'' \cite{Lind1999}
  equal to $\MOrder$.
\label{the:moesofts}
\end{Prop}

\begin{Cor}
  Given an exactly synchronizing \eM, the underlying sofic system is a
  subshift of finite type if and only if $\MOrder$ is finite.
\end{Cor}

\begin{Rem}
  A process with infinite Markov order can have a presentation whose
  underlying sofic system is a subshift of finite type.
\end{Rem}

These results draw out a connection with length scales of sofic
systems from symbolic dynamics \cite{Lind1999}. Subshifts of finite
type have a probability-agnostic length scale analog of the Markov
order known as the ``step''. In the case of \eM\ presentations, they
are in fact equal.

We will now prove that two of the lengths defined---the cryptic and
synchronization orders---are topological. That is, they are properties of the
presentation's graph topology and are independent of transition probabilities,
so long as changes to the probabilities do not remove transitions and do not
cause states to merge. Additionally, due to Prop.~\ref{the:moeso}, the Markov
order is topological. All three are topological since they depend only on the
length at which a conditional entropy vanishes, not on how it vanishes.

\begin{The}
  Synchronization order $\SOrder$ is a topological property of a
  presentation.
\label{the:soit}
\end{The}

\begin{ProThe}
  Beginning from Def.~\ref{def:so}, there is length $\ell = \SOrder$
  at which:
\begin{align*}
H[ \CausalState_\ell | \MeasSymbols{0}{\ell} ]
    = \sum_{w \in \MeasAlphabet^\ell}
    \Prob{(w)} H[ \CausalState_\ell | \MeasSymbols{0}{\ell} = w ] = 0 ~ .
\end{align*}
Thus, $H[\CausalState_\ell | \MeasSymbols{0}{\ell} = w] = 0$ for all
$w \in \MeasAlphabet^\ell$, the set of length-$\ell$ words with
positive probability. Since every word of length $\ell$ is
synchronizing, $\ell$ is certainly greater than the synchronization
order. As synchronizing words are synchronizing regardless of their
probability of occurring, the synchronization order $\SOrder$ is
topological. \qed
\end{ProThe}

\begin{Cor}
  Markov order $\MOrder$ is a topological property of an \eM.
\label{the:moit}
\end{Cor}

\begin{ProCor}
  Since $\SOrder$ is a topological property by Thm.~\ref{the:soit} and
  since an \eM's $\MOrder = \SOrder$ by Prop.~\ref{the:moeso}, the
  Markov order is topological. \qed
\end{ProCor}

\begin{The}
  Cryptic order $\COrder$ is a topological property of a presentation.
\label{the:coit}
\end{The}

\begin{ProThe}
  Beginning from Def.~\ref{eq:CrypticOrder}, there is a length $\ell =
  \COrder$ at which:
\begin{align*}
  0 & = H[ \CausalState_\ell | \Future ] \\
  & \stackrel{(1)}{=} \sum_{\mathclap{\future \in \MeasAlphabet^\infty}}
  \Prob{(\future)} H[ \CausalState_\ell | \Future = \future ] \\
  & \stackrel{(2)}{=} \sum_{\mathclap{w \in \Lsync}}
  \Prob{(w, \causalstate_w)} H[ \CausalState_\ell |
  \MeasSymbols{0}{|w|} = w, \CausalState_{|w|} = \causalstate_w ]
  ~.
\end{align*}
Here, step $(1)$ simply expands the conditional entropy. Step $(2)$ is
true provided that the sum is over minimal synchronizing words and
$\causalstate_w$ is the state to which one synchronizes via $w$. This
final sum is zero only if the sum vanishes term-by-term. Thus, given a
word that synchronizes and the state to which it synchronizes, each term
provides a \emph{cryptic-order candidate}---the number of states that
could not be retrodicted from that state and word. Finally, the
longest such cryptic order candidate is the cryptic order for the
presentation. \qed
\label{pro:coit}
\end{ProThe}

Restated, the cryptic order $\COrder$ is topological as it depends
only on the minimal synchronizing words, which are topological by
definition.

% No need to discuss transition function since it is implicitly used
% to find the minimal synchronizing words.

% it would be awesome to get the proof that if l_min is the length of the
% shortest msw, and l_max is the longest (possibly infinity) msw, then for all
% i in range(l_min, l_max), there exists a msw with length i.

\section{Algorithms}
\label{sec:algos}

We are now ready to turn to computing the various synchronization
length scales given a presentation. While all of the algorithms to
follow have compute times that are exponential in the number of
machine states, we find them to be very efficient in practice. This is
particularly the case when compared to naive algorithms to compute
these properties. For example, computing synchronization, Markov, or
cryptic orders by testing successively longer blocks of symbols is
exponential in the length of the longest block tested. Worse, in the
case of non-Markovian and $\infty$-cryptic processes the naive
algorithm will not halt. In addition, the naive implementation of
Thm.~\ref{the:coit} given in the proof to compute the cryptic order
has a compute time of $O(2^{2^N})$, whereas the one presented below is
a simple exponential of $N$.

Unsurprisingly, given the results provided in Sec.~\ref{sec:thes}, we
begin with the minimal synchronizing words as they are the
underpinnings of the synchronization and cryptic orders. The
algorithms make use of standard procedures. Most textbooks on
algorithms provide the necessary background; see, for example, Ref.
\cite{Cormen2009}.

\subsection{Minimal Synchronizing Words}

We construct a deterministic finite automaton (DFA) that recognizes
$\Lsync$ of a given presentation $\mathcal{M} = (Q, E)$, where $Q$ are
the states and $E$ are the edges. This is done as follows:
\begin{Alg}~
\begin{enumerate}
\setlength{\topsep}{0pt}
\setlength{\itemsep}{0pt}
\setlength{\parsep}{0pt}
\item Begin with the recurrent presentation $\mathcal{M}$.
\item Construct $\mathcal{M}$'s power automaton $2^{\mathcal{M}}$, producing
    a DFA $\mathcal{T} = 2^{\mathcal{M}}$.
\item Set the node in $\mathcal{T}$ that corresponds to all $\mathcal{M}$'s
    states as $\mathcal{T}$'s start state.
\item Remove all edges between singleton states of $\mathcal{T}$.
    (These are the edges from $\mathcal{M}$.)
\item Set all singleton states of $\mathcal{T}$ as accepting states.
\end{enumerate}
\label{algo:msw}
\end{Alg}

Now, we enumerate $\Lsync$ via an ordered breadth-first traversal of
$\mathcal{T}$, outputting each accepted word.

\subsection{Synchronization Order}
\label{algo:so}

Thanks to Eq.~(\ref{the:soit}) we see that $\SOrder$ is the shortest
length $\ell$ that encompasses all of $\Lsync$. This is, trivially,
the longest word in $\Lsync$. With this, computing the synchronization
order reduces to:
\begin{Alg}~
\begin{enumerate}
\setlength{\topsep}{0pt}
\setlength{\itemsep}{0pt}
\setlength{\parsep}{0pt}
\item If $\Lsync$ is infinite, return $\infty$.
\item Enumerate each word in $\Lsync$ and return the length of the longest word.
\end{enumerate}
\end{Alg}
The test in the first step can be done simply by running a
loop-detection algorithm on DFA $\mathcal{T}$. If there is a loop,
then $\Lsync$ is infinite.

\subsection{Markov Order}
\label{algo:mo}

Due to Thm.~\ref{the:moeso}, a process's Markov order can be computed
by finding the synchronization order of the process's \eM. If one does
not have the \eM\ for a process, but rather some other unifilar
presentation, it is still possible in some cases to obtain the Markov
order through the synchronization order. That is, the algorithms for
$\SOrder$ and $\COrder$ provide probes into the presentation's length
scales. It can be the case that $\MOrder$ is accessible to those
probes, if $\COrder < \SOrder$, but it is only guaranteed to be
accessible in the case of \eMs. Note, there exist techniques for
constructing the \eM\ from any presentation~\cite{Ellison2009}.

\subsection{Cryptic Order}
\label{algo:co}

In the following algorithm $\mathcal{T}$ refers to the power automaton of the
machine $\mathcal{M}$. $\mathcal{T}$'s states---$p$, $q$, and $r$---are
elements of the power set of the states of $\mathcal{M}$. By the
\emph{predecessors} of a state $q$ along edge $\edge{p}{x}{q}$ we refer to the
set $p' = \{ m | (\edge{m}{x}{n}) \in \mathcal{M} ~\mathrm{and}~ m \in p
~\mathrm{and}~ n \in q \}$. These are the states $m \in p$ that actually
transition to a state $n \in q$ on symbol $x$. By \emph{subset construction}
below we refer to the standard NFA-to-DFA conversion algorithm
\cite{Hopcroft2001}.

\begin{Alg}~
  \begin{enumerate}
    \setlength{\topsep}{0pt}
    \setlength{\itemsep}{0pt}
    \setlength{\parsep}{0pt}
  \item Construct the power automaton $\mathcal{T} = 2^{\mathcal{M}}$
    via subset construction.
  \item Push each edge $\edge{p}{x}{q}$ in $\mathcal{T}$ to a queue.
  \item While queue is not empty:
    \begin{enumerate}
      \setlength{\topsep}{0pt}
      \setlength{\itemsep}{0pt}
      \setlength{\parsep}{0pt}
    \item Pop edge $\edge{p}{x}{q}$ in the queue.
    \item If edge is in processed list:
      \begin{enumerate}
        \setlength{\topsep}{0pt}
        \setlength{\itemsep}{0pt}
        \setlength{\parsep}{0pt}
      \item Restart loop, popping the next edge from the queue.
      \end{enumerate}
    \item Find the predecessors $p'$ of $q$ along $\edge{p}{x}{q}$.
    \item If $p' \ne p$:
      \begin{enumerate}
        \setlength{\topsep}{0pt}
        \setlength{\itemsep}{0pt}
        \setlength{\parsep}{0pt}
      \item Remove edge $\edge{p}{x}{q}$ from $\mathcal{T}$.
      \end{enumerate}
    \item If $|p'| > 1$:
      \begin{enumerate}
        \setlength{\topsep}{0pt}
        \setlength{\itemsep}{0pt}
        \setlength{\parsep}{0pt}
      \item Perform subset construction on $p'$ (implicitly, this
        adds the edge $\edge{p'}{x}{q}$ to $\mathcal{T}$).
      \item Push each edge created in the prior step into the queue.
      \item For each $\edge{r}{y}{p}$ in $\mathcal{T}$:
        \begin{enumerate}
          \setlength{\topsep}{0pt}
          \setlength{\itemsep}{0pt}
          \setlength{\parsep}{0pt}
        \item Add edge $\edge{r}{y}{p'}$ to $\mathcal{T}$.
        \item Add edge $\edge{r}{y}{p'}$ to the queue.
        \end{enumerate}
      \item Add $\edge{p}{x}{q}$ and $\edge{p'}{x}{q}$ to the
        processed list.
      \end{enumerate}
    \end{enumerate}
  \end{enumerate}
\end{Alg}
The result is an automaton $\mathcal{T'}$. The longest path in
$\mathcal{T'}$ through transient states ending in a recurrent state is
the cryptic order. Skipping previously processed edges is important
since for some topologies the algorithm can enter a cycle where it
will remove and then later add the same edge, ad infinitum.

There are three simple additions to this algorithm that result in a
sizable decrease in running time. The first is to store the edges to
be processed in a priority queue, such that an edge $\edge{p}{x}{q}$
is popped before an edge $\edge{r}{y}{s}$ if $|q| < |s|$, or if $|q| =
|s|$, then $|p| < |r|$. The second optimization is to trim
dangling states after each pass through the outer loop. A dangling
state is a state $p$ such that there is no path from $p$ to the
recurrent states. The last method for improving speed is to not add
edges between recurrent states to the queue in step 2.

This algorithm for computing the cryptic order only holds for unifilar
presentations.

\section{Statistical Measures of Synchronization}

\subsection{The Synchronization Distribution}

Taking a slightly more general view than the synchronization order, we
consider statistical properties of synchronization, rather than just
the absolute length at which an ensemble will all be synchronized. In
this vein, we define a distribution that gives the probability for a
word to first synchronize at length $\ell$.

\begin{Def}
  The \emph{synchronization distribution} $S$ gives the probability of
  synchronizing to a presentation at length $\ell$:
\begin{align}
  S(\ell) \equiv \sum_{\mathclap{w \in \Lsync}}
  \Prob{(w)} \delta(|w| - \ell) ~.
\end{align}
where $\delta$ is the Kronecker delta function.
\label{def:sd}
\end{Def}

\begin{Rem}
  $S$ is normalized: $\sum_{\ell=0}^{\infty}{S(\ell)}=1$.
\end{Rem}

We now draw out two particular quantities from this
distribution---quantities that have observable meaning for a
presentation.

\begin{Def}
  The \emph{synchronization time} $\tau$~\cite{Crutchfield2010} is the
  average number of observations needed to synchronize to a
  presentation:
\begin{align}
  \tau & \equiv \sum_{ \mathclap{w \in \Lsync} } |w| \Prob{(w)} \\
  & = \mathbb{E}_{\ell}[S(\ell)] ~,
\end{align}
where the second equality shows that $\tau$ is also equal to the
expectation value of the synchronization distribution.
\label{def:st}
\end{Def}

The synchronization time is useful for understanding how long it takes \emph{on
average} to synchronize to a model. This is in contrast to the Markov order
which is the minimal longest-synchronization-time across all presentations of a
process.

\begin{Def}
  The \emph{synchronization entropy} $\SE$ is the uncertainty in the
  synchronization distribution:
\begin{align}
  \SE \equiv H[S(\ell)] ~.
\end{align}
% Hey now, for even, odd, and nemo processes $\SE = 2 + \EE$ \ldots
\label{def:se}
\end{Def}

\begin{Rem}
  Note that this is quite distinct from the \emph{synchronization
    information} $\SI$ of Ref.~\cite{Crutchfield2008}:
\begin{align*}
  \SI = \sum_{\ell = 1}^\infty H[\CausalState_\ell |
  \MeasSymbols{0}{\ell}] ~.
\end{align*}
\end{Rem}

The synchronization entropy, in contrast, measures the flatness of the
synchronization distribution. And, since the synchronization
distribution decays exponentially with length, the fatter the tail,
the higher the uncertainty in synchronization.

\subsection{The Synchronization Distribution}
\label{algo:sd}

There are two methods to compute the synchronization distribution. The
first requires an \eM\ with finite recurrent and transient components.
The second, only a finite recurrent component. We present the former
case first.

\begin{Alg}~
  \begin{enumerate}
    \setlength{\topsep}{0pt}
    \setlength{\itemsep}{0pt}
    \setlength{\parsep}{0pt}
  \item Perform an ordered breadth-first traversal of the finite \eM.
  \item While traversing, keep track of the word induced by the path
    and the product of the probabilities along that path.
  \item When a recurrent node is reached, stop that particular thread
    of the traversal.
  \item Sum the probabilities of all words with the same length.
  \end{enumerate}
\end{Alg}
This algorithm produces each minimal synchronizing word and its
probability in lexicographic order. Then words of each length can be
grouped and their probabilities summed to get the synchronization
distribution.

The second algorithm is used when an \eM\ with finite transient
structure is not available:
\begin{Alg}~
  \begin{enumerate}
    \setlength{\topsep}{0pt}
    \setlength{\itemsep}{0pt}
    \setlength{\parsep}{0pt}
  \item Produce all the minimal synchronizing words from the DFA given
    in Algorithm~\ref{algo:msw}.
  \item For each minimal synchronizing word, compute its probability
    using the recurrent \eM~\cite{Crutchfield2008}.
  \item Sum the probabilities of all words with the same length.
  \end{enumerate}
\end{Alg}

\begin{figure}[ht]
  \centering
  \includegraphics[width=0.9\columnwidth]{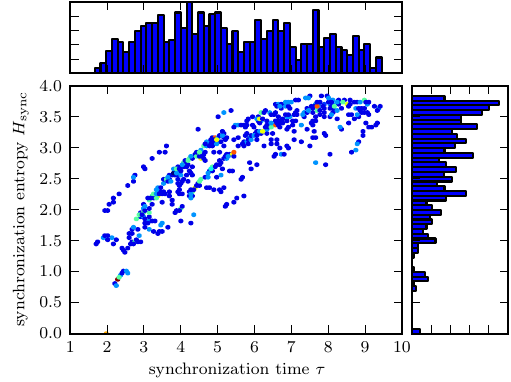}
  \caption{Distribution of synchronization time $\tau$ and
    synchronization entropy $\SE$ for all $1,388$ four-state,
    binary-alphabet, exactly-synchronizing \eMs\ with uniform outgoing
    transition probabilities. Individual histograms for each property
    are shown above and to the right. }
\label{fig:hists}
\end{figure}

Once the distribution is computed using one of the above algorithms,
it is trivial to compute the Shannon entropy and mean of the
distribution to get the synchronization entropy and synchronization
time, respectively.

\subsection{Results}
\label{appendix:results}

Finally, we survey the distribution of synchronization times $\tau$
and synchronization entropies $\SE$ for all $1,388$ four-state,
binary-alphabet, exact \eMs\ with uniform outgoing transition
probabilities \cite{Johnson2010}. See Fig.~\ref{fig:hists}. It is
interesting to note that there is structure in the distribution in the
form of veils. However, the veils are not the entirety of the
distribution, there are many machines that fall elsewhere.

\end{document}